\documentclass[a4paper,11pt]{article}
\pdfoutput=1 

\usepackage{jcappub} 

\usepackage[T1]{fontenc} 

\usepackage{newtxtext,newtxmath}

\title{\boldmath Constraints on multi-fluid cosmology in $f(G)$ gravity models with different observational data sets}

\author[a]{Praveen Kumar Dhankar}
\author[b,c,1]{Albert Munyeshyaka,\note{Corresponding author.}}
\author[d]{Aritra Sanyal}
\author[e]{Safiqul Islam}
\author[d]{Farook Rahaman}

\affiliation[a]{Symbiosis Institute of Technology, Nagpur Campus, Symbiosis International (Deemed University), Pune-440008, Maharashtra, India}
\affiliation[b]{Rwanda Astrophysics Space and Climate Science Research Group, University of Rwanda, College of Science and Technology, Kigali, Rwanda}
\affiliation[c]{Kibogora Polytechnic, Faculty of education, Western province, Rwanda}

\affiliation[d]{Department of Mathematics, Jadavpur University, Kolkata 700032, West Bengal, India}

\affiliation[e]{Department of Basic Sciences, General Administration of Preparatory Year, King Faisal University, P.O. Box 400, Al Ahsa 31982, Saudi Arabia;\\Department of Mathematics and Statistics, College of Science, King Faisal University, P.O. Box 400, Al Ahsa 31982, Saudi Arabia}

\emailAdd{pkumar6743@gmail.com}
\emailAdd{munalph@gmail.com}
\emailAdd{aritrasanyal1@gmail.com}
\emailAdd{sislam@kfu.edu.sa}
\emailAdd{rahaman@associates.iucaa.in}

\abstract{
 In the present work, we incorporate redshift space distortion measurement to investigate the growth of large scale structure within the framework of multi-fluid cosmology in the context of $f(G)$ gravity. Using 3 different $f(G)$ gravity models, where $f(G)$ is the function of Gauss-Bonnet invariant, we compare the predictions of $f(G)$ gravity expansion history--through the Friedmann equation with Hubble data, BAO data sets and constrain models parameters such as $\Omega_{m}$ and $H_{0}$. Within the context of multi-fluid cosmology in $f(G)$ gravity, we obtain the structure growth equation. This equation is then combined with $\sigma_{8}$ to get $f\sigma_{8}$ predictions--which is compared with redshift space distortion data to constrain models parameters to obtain best-fit values including $\sigma_{8}$. This involves performing a Markov Chain Monte Carlo (MCMC) analysis for these specific forms of $f(G)$ models. }

\begin{document}
\maketitle
\flushbottom

\section{Introduction}\label{introduction}
Recent cosmic acceleration \cite{hogg2005cosmic,lu2007obtaining} confirmed by the set of accurate data from cosmological observations at both background and perturbations leads to considering modified theories of gravity as alternative theory to explain the dynamics of the universe without the need for dark energy hypothesis \cite{copeland2006dynamics,bertschinger2008distinguishing,nojiri2007introduction,riess1998observational}. \\
Among the most considered modified theories of gravity  include the $f(R)$, $f(T)$, $f(Q)$ and $f(G$ gravity theories, where $R$, $T$, $Q$ and $G$ are the Ricci scalar, torsion scalar, non-metricity scalar and Gauss-Bonnet invariant, respectively. The most challenging issue in each modified theory of gravity is perhaps to determine the viable functional form of the model. Even though some general analysis such as the existence of Noether symmetries, the absence of ghosts, the stability of perturbations can be extracted through theoretical arguments, there is a need to confront theoretical model with observations \cite{anagnostopoulos2019bayesian}.\\

In this context, different observation works have been conducted using solar system data \cite{iorio2012solar} and cosmological data \cite{capozziello2015transition,bonici2019constraints,pan2024interacting}. Furthermore, the confrontation with cosmological data used mainly Supernovae type $I_{a}$ data, Cosmic Microwave Background (CMB), Baryonic Acoustic Oscillations (BAO), and
Hubble data observations to constrain background evolution. In the work conducted by Anagnostopoulos F.K \cite{anagnostopoulos2019bayesian}, the authors extend the analysis to the perturbation level using the $f\sigma_{8}$ data to constrain the $f(T)$ gravity.\\

In addition, the modified theory of $f(Q)$ gravity when limited to observational data leads to interesting cosmological phenomenology at the background level \cite{yang2024data,sahlu2024constraining,mandal2020cosmography}. In the work carried out by \cite{enkhili2024cosmological}, the authors constrained cosmological dynamical dark energy model in the context of $f(Q)$ gravity using Markov Chain Monte Carlo (MCMC) analysis. In addition, different authors have been testing this kind of theory against various observational data, which include both background and perturbation observations \cite{sahlu2024constraining,mhamdi2024cosmological,anagnostopoulos2021first,barros2020testing,lazkoz2019observational}, and revealed that $f(Q)$ gravity can challenge the standard $\Lambda$CDM scenario. In \cite{barros2020testing}, the authors tested $f(Q)$ gravity with redshift space distortions and confirmed that the best fit parameters reveal that the $\sigma_{8}$ tension between Plank and large scale structure data can be alleviated within the considered framework.  In \cite{lazkoz2019observational}, the authors constrained the $f(Q)$ gravity using different observational probes and confirmed that the approach used can provide a different perspective on the formulation of observationally reliable alternative model of gravity. The work carried out in the works presented in \cite{mhamdi2024constraints,mhamdi2024cosmological,sahlu2025structure,sahlu2024constraining} treated $f(Q)$ gravity using the mentioned observational data and the constrained model parameters.\\  The authors in Ref. \cite{nojiri2005modified} suggested the modified gravity where arbitrary function of Gauss-Bonnet term is added to the Einstein action as gravitational dark energy and showed that this theory may pass solar system tests and may describe the most interesting features of late-time cosmology. In Ref. \cite{lohakare2024cosmology}, the authors presented a method for numerically solving the Friedmann equations of the modified Gauss-Bonnet $f(G)$ gravity in the presence of pressureless matter and applied the Bayesian MCMC technique using late-time cosmic observations to impose limitations on free parameters of the GAuss-Bonnet model. The authors showed that the $f(G)$ model can reproduce the low redshift behavior of the $\Lambda$CDM model.

 Additionally, studying cosmological perturbations helps to analyse the dynamics of the Universe, including large scale structure formation for both General relativity and modified theories of gravity. the work considered in \cite{munyeshyaka2023multifluid} analysed multifluid cosmology in $f(G)$ gravity and presented numerical results of energy density contrast ($\delta(z)$) on both long and short wavelength limits and showed that the energy density contrast decays with increase in redshift. Motivated by the mentioned works, In the present work, we will use the $1+3$ covariant and Gauge-Invariant formalism \cite{challinor2000microwave,dunsby1992cosmological,dunsby1991gauge,ellis1989covariant,ellis2011inhomogeneity,hawking1966perturbations,ellis1989covariant,abebe2012covariant, abebe2015breaking,ntahompagaze2017f, ntahompagaze2018study, carloni2006gauge,sami2021covariant,li2007cosmology,munyeshyaka2021cosmological} instead of metric formalism \cite{de2008evolution,bardeen1980gauge,kodama1984cosmological,bertschinger2000cosmological,dunsby1992cosmological,dunsby1991gauge}, following the work conducted by Munyeshyaka et al. \cite{munyeshyaka2023multifluid} .  In doing so, we will consider the approximated equation using the quasi-static approximation for the dust-dominated case in the context of $f(G)$ gravity and obtain structure growth equation.  \\For pedagogical purpose, we incorporate $3$ different $f(G)$ gravity models, which includes the general $f(G)$ model given by $f(G)=G-\frac{1}{2}\Big(\sqrt{\frac{6m(m-1)G}{(m+1)^{2}}}+AG^{\frac{3}{4}m(1+w)}\Big)$ dubbed (model A), power-law $f(G)$ model given by $f(G) = \alpha G^\beta$, dubbed (model B) and an exponential f(G) model given by $f(G) = \alpha G_0(1 - e^{-pG/G_0})$, dubbed (model C) functions of Gauss-Bonnet invariant $G$, where $m$, $A$, $w$, $\alpha$, $\beta$, $G_{0}$, and $p$ are constants.\\
After obtaining the modified Friedmann equation and the structure grow equation, we employ thorough observational analysis at both background and perturbation levels in the context of modified Gauss-Bonnet gravity. Therefore, the aim of our study is to constrain $f(G)$ gravity models by means of Bayesian analysis Hubble measurement, BAO measurement and the redshift space distortion data sets. \\In order to achieve this aim, we initially solve the modified Friedmann equations under the assumption that the Universe is exclusively composed of dust matter, where the equation of state parameter vanishes . We rigorously evaluate the performance of each model by assessing its accuracy at both the background and perturbation levels. To estimate model parameters.\\
The next task is to use MCMC analysis to constrain model parameters resulting from the use of $f(G)$ gravity models. In so doing, we compute the corner (triangular) plots and present the mean value parameters corresponding to the combinations of data sets namely: (i) H(z) measurements (ii) BAO measurements. For further analysis, we use joint analysis of iii) H(z)+BAO data and iv) the  redshift space distortion data $f\sigma_{8}$ data with the latest measurements of the growth rate and amplitude of matter fluctuations $\sigma_{8}$ namely the $\sigma_{8}+f+f\sigma_{8}$. Using the MCMC simulations, we are able to constrain the best fit model parameters in the context of $f(G)$ gravity theory for each model.\\\\
The next part of this paper is organised as follows: in Section (\ref{sec2}), we describe the $f(G)$ gravity theory and cosmology, where the background and perturbation equations are presented. In this section, we obtained the modified Friedmann equation and structure grow equation and presented different $f(G)$ models under consideration. In Section (\ref{sec3}), we describe the data and method used to achieve our aim. Section (\ref{sec4}) presents and discusses the results, whereas
Sections (\ref{sec5}) is reserved for conclusion.\\
The adopted spacetime signature is $(-,+,+,+)$ and unless stated otherwise, we
use $\mu$, $\nu$ . . . $= 0$, $1$, $2$, $3$, and $8\pi G_{ N} = c = 1$, where $G_{ N}$ is the gravitational constant
and $c$ is the speed of light, and we consider the Friedmann-Robertson-Walker (FRW) spacetime background in this work. The symbols $\bigtriangledown$ represent the usual covariant derivative, and
$\partial$ corresponds to partial differentiation, and an over-dot represents differentiation with
respect to proper time.
\section{The modified Gauss-Bonnet $f(G)$ gravity and cosmology}\label{sec2}
In this section, we review the cosmological equations in the framework of multi-fluid $f(G)$ gravity theory. In this regard, the action for an arbitrary function of Gauss-Bonnet gravity is rewritten as

\begin{eqnarray}
 S=\frac{1}{2\kappa^{2}}\int d^{4}x \sqrt{-g}\left( R+f(G)+\mathcal{L}_{m} \right),
 \label{eq01}
\end{eqnarray}
where $ \kappa = 8\pi G_{N}$ is a constant, $G_{N}$ is the Newton gravitational constant, $f(G)$ is a differentiable function of the Gauss-Bonnet invariant $G$ and $\mathcal{L}_{m}$ is the matter Lagrangian \cite{kawai1999evolution,nojiri2011unified,satoh2008circular,satoh2008higher,kawai1998instability,sberna2017nonsingular,odintsov2020rectifying,nojiri2019ghost, zheng2011growth, oikonomou2015singular}.
The Gauss-Bonnet term is given by
 \begin{equation}
  G=R^{2}-4R_{\mu \nu}R^{\mu\nu}+R_{\mu\nu \sigma \lambda}R^{\mu\nu \sigma \lambda},
 \end{equation}
where  $R$, $R_{\mu \nu}$ and $R_{\mu\nu \sigma \lambda}$ are the Ricci
scalar, Ricci tensor and Riemann tensor. respectively. The information about the  content of the universe
is contained within the energy-momentum tensor $ T_{\mu \nu} $.
 By varying the action (eq. \ref{eq01}) with respect to the metric $g_{\mu\nu}$,  the modified Einstein equation is given by
  \begin{eqnarray}
   && R_{\mu\nu}-\frac{1}{2}g^{\mu\nu}R= T^{m}_{\mu\nu}+\frac{1}{2}g^{\mu\nu}f-2f'RR^{\mu\nu}+4f'R^{\mu}_{\lambda}R^{\nu\lambda}-2f'R^{\mu\nu\sigma\tau}R^{\lambda\sigma\tau}_{b}\nonumber\\
   && \quad  \quad -4f'R^{\mu\lambda\sigma\nu}R_{\lambda\sigma}+2R\bigtriangledown^{\mu}\bigtriangledown_{\nu}f'-2Rg^{\mu\nu}\bigtriangledown^{2}f'-4R^{\nu\lambda}\bigtriangledown_{\lambda}\bigtriangledown^{\mu}f'\nonumber\\
   && \quad \quad-4R^{\mu\lambda}\bigtriangledown_{\lambda}\bigtriangledown^{\nu}f'+4R^{\mu\nu}\bigtriangledown^{2}f' +4g^{\mu\nu}R^{\lambda\sigma}\bigtriangledown_{\lambda}\bigtriangledown_{\sigma}f'-4R^{\mu\lambda\nu\sigma}\bigtriangledown_{\lambda}\bigtriangledown_{\sigma}f'\;,
   \label{eq3}
  \end{eqnarray}
where $f\equiv f(G)$ and $f'=\frac{\partial f}{\partial G}$. $T^{m}_{\mu\nu}$ is the energy momentum tensor of the matter fluid (photons,baryons, cold dark matter, and light neutrinos). The equation eq. (\ref{eq3}) can be rewritten in compact form as
\begin{equation}
 R_{\mu\nu}-\frac{1}{2}g^{\mu\nu}R= T^{m}_{\mu\nu}+T_{\mu\nu}^{G},
 \label{eq4}
\end{equation} where
\begin{eqnarray}
 &&T_{\mu\nu}^{G}=\frac{1}{2}g^{\mu\nu}f-2f'RR^{\mu\nu}+4f'R^{\mu}_{\lambda}R^{\nu\lambda}-2f'R^{\mu\nu\sigma\tau}R^{\lambda\sigma\tau}_{b}-4f'R^{\mu\lambda\sigma\nu}R_{\lambda\sigma}\nonumber \\
 && \quad \quad+2R\bigtriangledown^{\mu}\bigtriangledown_{\nu}f'
 -2Rg^{\mu\nu}\bigtriangledown^{2}f'-4R^{\nu\lambda}\bigtriangledown_{\lambda}\bigtriangledown^{\mu}f'-4R^{\mu\lambda}\bigtriangledown_{\lambda}\bigtriangledown^{\nu}f'\nonumber\\
 &&\quad\quad+4R^{\mu\nu}\bigtriangledown^{2}f'+4g^{\mu\nu}R^{\lambda\sigma}\bigtriangledown_{\lambda}\bigtriangledown_{\sigma}f'-4R^{\mu\lambda\nu\sigma}\bigtriangledown_{\lambda}\bigtriangledown_{\sigma}f'\;
 \label{eq5}
\end{eqnarray} is the energy-momentum tensor resulting from the Gauss-Bonnet contribution.
For the case $f(G)=G$, equation eq. (\ref{eq5}) vanishes  ($T_{\mu\nu}^{G}=0)$, hence Einstein gravity is recovered ($R_{\mu\nu}-\frac{1}{2}g^{\mu\nu}R= T^{m}_{\mu\nu}$).
In the following section, we present the modified Friedmann equation which will help us to constrain parameters for certain $f(G)$ models.
\subsection{Background behavior}
In order to proceed to the cosmological application of $f(G)$ gravity, one imposes the homogeneous and isotropic flat Friedmann-Robertson-Walker (FRW) geometry, so that the metric is given by
\begin{equation}
 ds^{2}=-dt^{2}+a^{2}dX^{2},
\end{equation}
 where $X=x,y,z$, and $a(t)$ is the cosmological scale factor. Inserting this choice into the energy-momentum tensor, the equation (0,0 component of eq.( \ref{eq3})) corresponding to the Friedmann equation for $f(G)$ gravity is presented as
\begin{equation}
 3H^{2}=\frac{1}{2}\left( Gf'-f-24\dot{G}H^{3}f''\right)+\rho_{m}\;.
 \label{eq2.7}
 \end{equation}  The Gauss-Bonnet and Ricci scalars are given as $G=24H^{2}(\dot{H}+H^{2})$ and $R=6(\dot{H}+2H^{2})$, respectively, and $H=\frac{\dot{a}}{a}$ is the Hubble parameter. For the case of multi-fluid $f(G)$ gravity, the total energy density and total pressure are modified as
$\rho_{total}=3H^{2}=\rho_{total}=\rho_{G}+\rho_{m} ~~ p_{total}=-(3H^{2}+2\dot{H})=p_{G}+p_{m}$, respectively.
   The Gauss-Bonnet energy density and pressure representing dark energy terms are given by \cite{garcia2011energy}
\begin{eqnarray}
&& \rho_{G}=\frac{1}{2} \Big(Gf'-f-24\dot{G}H^{3}f''\Big)\;,\\
&& p_{G}=\frac{1}{2}\Big( f-f'G+\frac{2G\dot{G}}{3H}f''+8H^{2}\ddot{G}f''+8H^{2}\dot{G}^{2}f'''\Big).
\end{eqnarray}
For the case $f(G)=G$, $\rho_{G}=0=p_{G}$.
We assume that matter can be described by a barotropic perfect fluid such that $p=w\rho$. In the FRW universe, the energy conservation law can be expressed as a standard equation.
\begin{equation}
 \dot{\rho}+3H(\rho+p)=0.
\end{equation}
 The general solution is given by
\begin{equation}
 \rho=\rho_{0}t^{-3m(1+w)}.
 \label{eq2.11}
\end{equation}
Eq. (\ref{eq2.7}) is crucial to constraining the model parameters. This will be done in the next sections after defining different pedagogical $f(G)$ models. The next aim of this work is to study the effects of multi-fluid cosmology on the structure growth in the context of $f(G)$ gravity and to constrain the $\sigma_{8}$ parameter. In order to achieve our aim, we need to present the perturbation equation on small scale obtained by considering the quasi-static approximation. After getting the perturbation equation, we therefore need to find the growth equation.
\subsection{Linear matter perturbation and structure growth equation}
It is now a well-established fact that the universe is not perfectly smooth but is full of large-scale structures such as galaxies, clusters, superclusters, voids, etc., seeded from primordial fluctuations. Cosmological perturbation theory provides the mechanism to explain how these small fluctuations grow and form the large-scale structures in the real, lumpy, universe. Following the work conducted by Munyeshyaka A et al. \cite{munyeshyaka2023multifluid}, the covariant formalism was used and perturbation equations were obtained. In this work we use the obtained perturbation equation in the  quasi-static approximation limit and represent it here  with some modification as
\begin{eqnarray}
 &&(1+z)^{2} H^{2} \Delta''_{d}=-(1+z)H
\{\frac{9}{4\theta^{3}}+\frac{1}{\theta}\rho_{d}
  +\frac{G}{2\theta}f'+f''(\frac{8\theta^{2}G'}{9}-\frac{3GG'}{\theta^{2}}-\frac{4}{3}\theta G'') \nonumber\\&&-\frac{1}{2\theta}f-\frac{2}{3}\theta
  -\frac{4}{3}\theta G'^{2}f''' +(1+z)H'\}\Delta'_{d}+\rho_{d}\Delta_{d}.
 \label{eq2.12}
\end{eqnarray} This is the energy over-density equation for a dust dominated universe in the context of modified Gauss-Bonnet gravity,
whereas for the case of $f(G)=G$, the energy over-density for the $\Lambda$CDM is presented as
\begin{eqnarray}
 \Delta''_{d}-\frac{2}{1+z}\Delta'_{d}-\frac{3}{(1+z)^{2}}\Delta_{d}=0.
 \label{eq2.13}
\end{eqnarray}
In order to constrain the model parameters for $f(G)$ gravity, we first need to define the structure growth equation and compute the combinations $y\sigma_{8} $. First, we define the linear growth rate $y$ as
\begin{equation}
y=-(1+z)\frac{\Delta'_{m}(z)}{\Delta_{m}(z_{in})}, \label{eq2.14}
\end{equation}
 where $\sigma_{ 8}$ is the root mean square normalisations of the matter power spectrum within the radius sphere of $8h^{-1} MPC$ \cite{sahlu2025structure,sahlu2024constraining}, given by
 \begin{eqnarray}
 \sigma_{8} (z)=\sigma_{8} (z_{in})\frac{\Delta_{m}(z)}{\Delta_{m}(z_{in})}.
\end{eqnarray} with $z_{in}$ is the initial redshift. Therefore the combination $y\sigma_{8} (z)$ results in \cite{kazantzidis2018evolution,kazantzidis2021sigma,nesseris2017tension,sahlu2025structure,panotopoulos2021growth}
\begin{eqnarray}
y\sigma_{8}  (z)=-(1+z)\sigma_{8}  (z)\frac{\Delta'_{m}(z)}{\Delta_{m}(z_{in})}. \label{eq2.16}
\end{eqnarray}
Using eq. (\ref{eq2.14}) in eq. (\ref{eq2.12}) and eq. (\ref{eq2.13}) we obtain
\begin{eqnarray}
 &&-(1+z)y'+y^{2}=-H^{2}y+\frac{1}{H}
\{\frac{1}{12H^{3}}+\frac{\Omega_{m}(1+z)^{3}}{3H}
  +\frac{G}{6H}f'+f''(8H^{2}G'-\frac{GG'}{3H^{2}}-4H G'') \nonumber\\&&-\frac{f}{6H}-2H
  -4H G'^{2}f''' +(1+z)H'\}y+3\frac{\Omega_{m}(1+z)^{3}}{4H^{2}}.
 \label{eq2.17}
\end{eqnarray}  Eq. (\ref{eq2.17}) is the theoretical growth equation in the context of $f(G)$ gravity. We will use this model to fit with observational data for constraining model parameters including $\sigma_{8}$. For the case of $\Lambda$CDM, we have
\begin{eqnarray}
 -(1+z)y'+3y+y^{2}-\frac{3\Omega_{m}(1+z)^{3}}{4H^{2}}=0.
 \label{eq2.18}
\end{eqnarray} Solving eq. (\ref{eq2.12}) for $\Delta(z)$ and eq. (\ref{eq2.17}) for $y$ allows to constrain model parameters using eq. (\ref{eq2.16}). In order to solve the Friedmann equation for the background and the structure growth equation, we need to define $f(G)$ model. For pedagogical and comparison purposes, we define three different $f(G)$ models as follows:
We close this section by presenting three specific viable
$f (G ) $ models namely: General $f(G)$ model (dubbed model A), power-law  and exponential $f(G)$ models dubbed model B and C, respectively,  that are efficient in successfully passing the confrontation with observations.
\begin{enumerate}
\item The general $f(G)$ gravity model (model A) is given by \cite{cognola2006dark,goheer2009coexistence,rastkar2012phantom, munyeshyaka2023multifluid}
$f(G)=G-\frac{1}{2}\Big(\sqrt{\frac{6m(m-1)G}{(m+1)^{2}}}+AG^{\frac{3}{4}m(1+w)}\Big)$, where $A=\frac{8\rho_{0}(m-1)\left[13824m^{9}(m-1)^{3}\right]^{-\frac{1}{4}m(1+w)}}{4+m\left[3m(1+w)(w+\frac{4}{3})-18w-19\right]}$
For the case $m=1$, $f(G)\sim G$,  In order to produce  an accelerating universe,  $m\succ 1$ and $4+m\Big[3m(1+w)(w+\frac{4}{3})-18w-19\Big]\neq 0$, with $w=[-1,1]$.
 \item  The power-law model (model B) is given by $f(G)=\alpha G^{\beta}$. For the case $\alpha=1$ and $\beta=1$, the $\Lambda$ CDM case is retained. This model resembles the one considered in \cite{lee2020viable}, for the case $a_{1}=\alpha$ and $b_{1}=0$.
 \item Motivated by exponential $f(R)$ \cite{linder2009exponential} gravity and $f(T)$ gravity \cite{nesseris2013viable} models, we can construct the exponential $f(G)$ gravity model (hereafter model C) as $f(G)=\alpha G_{0}\Big(1-e^{-p\frac{G}{G_{0}}}\Big)$, with $\alpha$, $G)_{0}$ and $p$ are model parameters. This reduces to $\Lambda$CDM for high values of $p$.
 \end{enumerate}
 In this work, we are motivated in the viability of $f(G)$ gravity models in the sense that these defined $f(G)$ models can i) describe the matter and dark energy eras ii) they are consistent with observational data iii) pass the solar system tests and iv) they have stable perturbations.  Although these important studies have not yet been performed for all defined $f(G)$ models, failure of a particular model to pass one of these tests is enough to be ruled out. As shown above, for all these $3$ $f(G)$ models, the distortion parameters measure the smooth deviation from $\Lambda$CDM. In the next section, we focus on them and apply them to the observational data for model parameters estimation.
\section{Data and methods}\label{sec3}
In this section, we perform a detailed observational investigation of three distinct $f(G)$ gravity models within the context of multi-fluid cosmology. The analysis is driven by several complementary cosmological datasets that probe both the background evolution of the universe and the growth of cosmic structures. Specifically, we utilise: (i) direct measurements of the Hubble parameter $H(z)$, which trace the expansion rate across redshift; (ii) Baryon Acoustic Oscillation (BAO) observables, which encode angular and radial distances from galaxy clustering surveys; and (iii) redshift-space distortion (RSD) data in terms of $f\sigma_8(z)$, alongside direct measurements of $f(z)$ and $\sigma_8(z)$ that constrain the growth of matter perturbations.
\begin{itemize}
	\item \textbf{CC Data:} This data set contains 32 model-independent observational points corresponding to the Hubble parameter, commonly referred to as Cosmic Chronometers (CC) \cite{Jimenez:2001gg,Moresco:2015cya}.
	
		\item \textbf{DESI BAO:} This data set includes samples of Baryon Acoustic Oscillations (BAO) from the Dark Energy Spectroscopic Instrument (DESI) Release II \cite{DESI:2025zgx}. The observables are \(\{D_M/r_d, D_H/r_d, D_V/r_d\}\), where \(D_M\) denotes the co-moving angular diameter distance, \(D_H\) the Hubble distance, \(D_V\) the spherically averaged distance, and \(r_d\) the sound horizon at the drag epoch, corresponding to the redshift \(z_d = 1060.0\) \cite{eBOSS:2020yzd,DESI:2024mwx,Planck:2018vyg},
		\begin{equation}
			r_{d} =\int_{ z_{d}}^{\infty} \dfrac{3 \times 10^{5} d{z}}{H \sqrt{3\left(1 + \frac{3 \Omega_{b_0}h^2}{4 \Omega_{\gamma_0}h^2(1+{z})}\right)}} \ .
			\label{sound_distance}
		\end{equation}
		Here,  \(\Omega_{\gamma_0} h^2\) denotes the photon density parameter and \(h \equiv H_0/100\), with the value \(2.472 \times 10^{-5}\) and $\Omega_{b_0}$ represents the baryon density at the present epoch with the value $\Omega_{b_0}h^2 = 0.02236$ \cite{Planck:2018vyg,Chen:2018dbv}. However, for this study we will treat \(r_d\) as a free parameter. We refer to this data set as `BAO'.
		
		\item \textbf{\boldmath $f\sigma_8$ Data :} This data sets includes $30$ observational samples of redshift-space distortion $f\sigma_8$ in the redshift \(z \in [0.001,1.944]\) \cite{sahlu2025structure}. We label this data set as `fs8'.

\end{itemize}

For the above data sets, we compute the joint likelihood corresponding to the following combinations of datasets: (i) CC+BAO  and ii) $\sigma_{8}$+f+fs8. This combination is applied to each $f(G)$ model. The joined likelihood is estimated as
\begin{equation}
	-2 \ln \mathcal{L}_{\rm tot} = \chi^2_{\rm tot}.
\end{equation}
 The likelihood is estimated by implementing the model in \texttt{Python} using the publicly available affine-invariant Markov Chain Monte Carlo (MCMC) ensemble sampler \texttt{emcee} \cite{Foreman-Mackey:2012any}. The resulting posterior distributions are visualized using triangular (corner) plots, generated by analysing the MCMC chains with \texttt{GetDist} \cite{Lewis:2019xzd}.

\section{Results and discussions}\label{sec4}

\begin{figure}[t!]
    \centering

        \includegraphics[height=60mm]{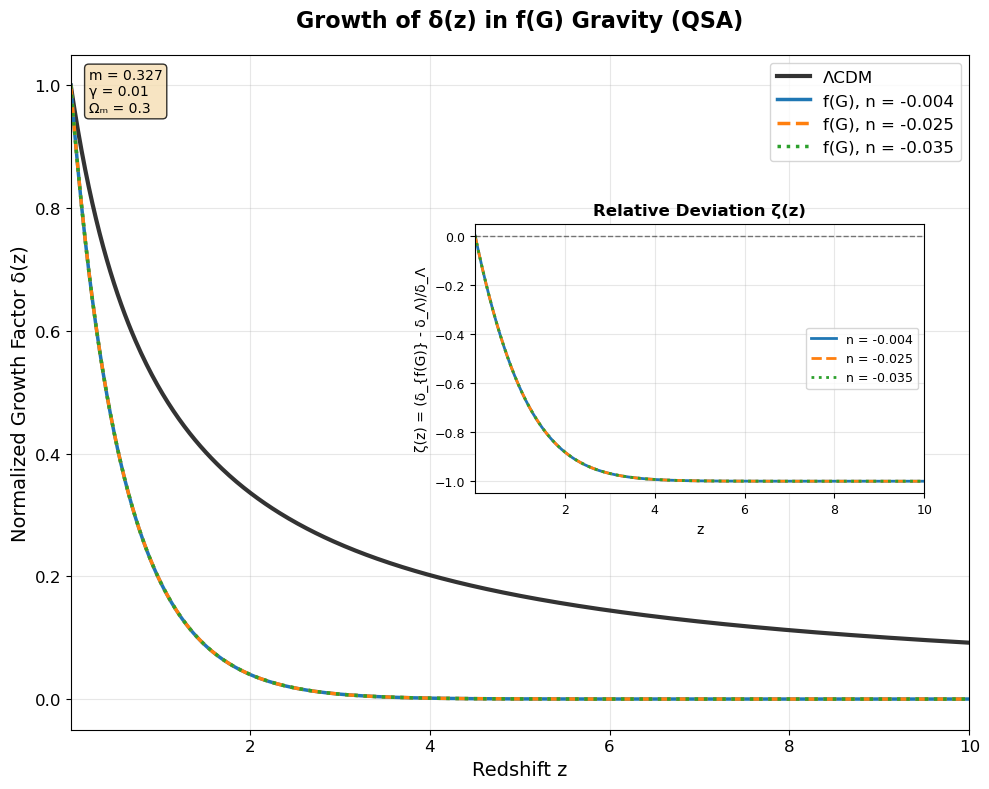}
       \caption{Growth of matter perturbations \(\delta(z)\) using Model A ($f(G)$ gravity) in the quasi-static approximation for different values of the model parameter \(m\). The inset shows the relative deviation from the \(\Lambda\)CDM model, highlighting the sensitivity of growth predictions to variations in \(m\). (Here $n$ shown in the plot is the same as $m$)}

        \label{fig1}
\end{figure}

\begin{figure}[t!]
    \centering

        \includegraphics[height=60mm]{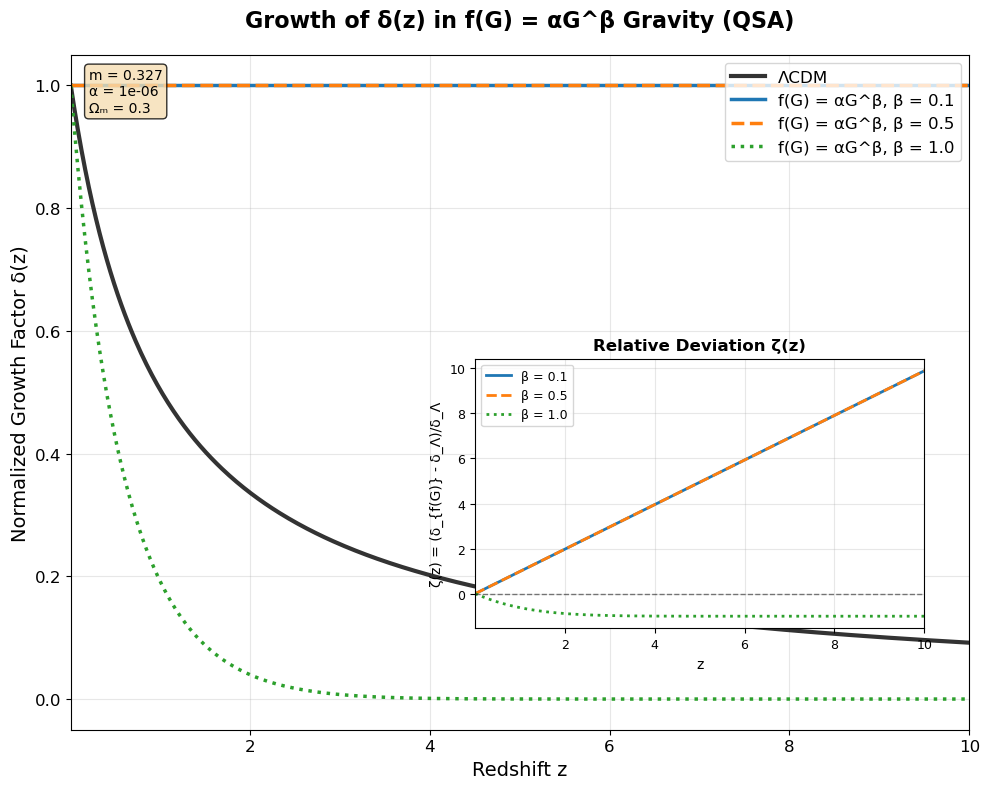}
       \caption{Growth of matter perturbations \(\delta(z)\) using Model B ($f(G)$ gravity) in the quasi-static approximation for different values of the model parameter ($\beta$). The inset shows the relative deviation from the \(\Lambda\)CDM model, highlighting the sensitivity of growth predictions to variations in $\beta$.}

        \label{fig2}
\end{figure}

\begin{figure}[t!]
    \centering

        \includegraphics[height=60mm]{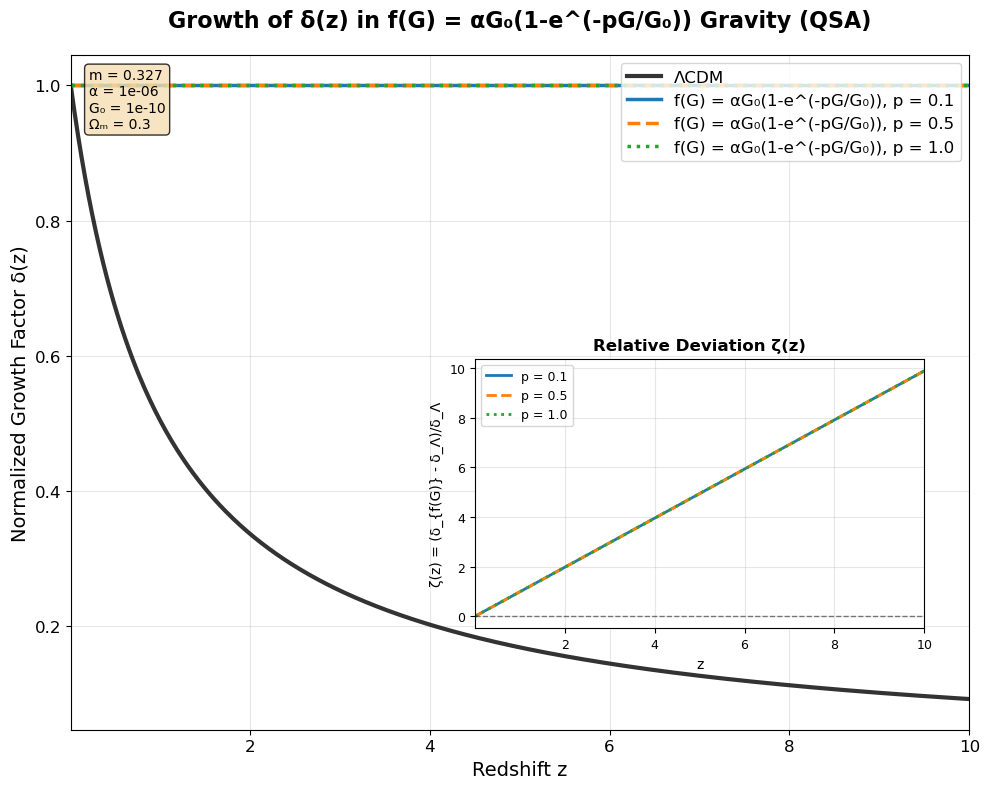}
       \caption{Growth of matter perturbations \(\delta(z)\) using Model C ($f(G)$ gravity) in the quasi-static approximation for different values of the model parameter \(p\). The inset shows the relative deviation from the ($\Lambda$)CDM model, highlighting the sensitivity of growth predictions to variations in \(p\).}

        \label{fig3}
\end{figure}

\begin{figure}
	\centering
	(a)\includegraphics[scale=0.3]{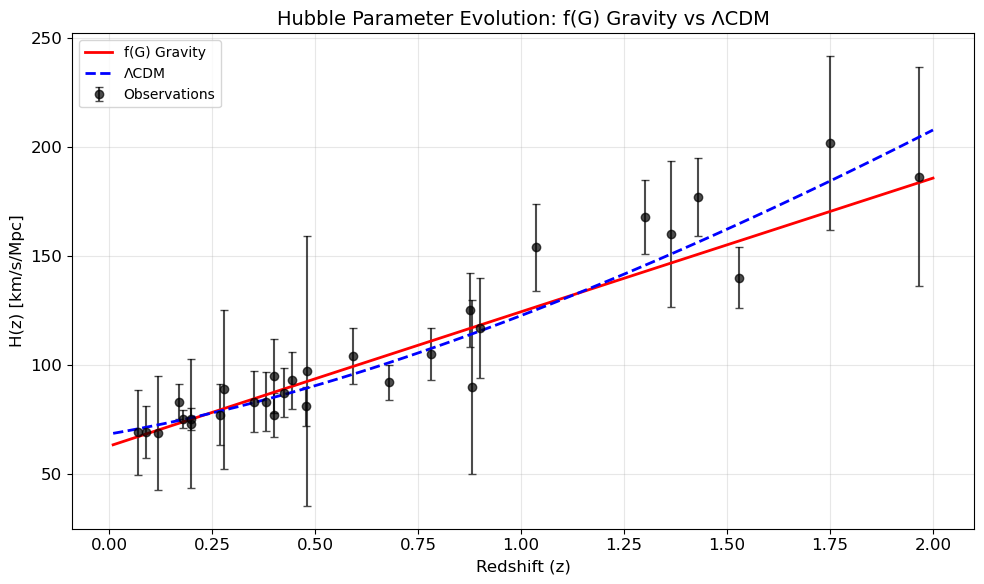}
	(b)\includegraphics[scale=0.3]{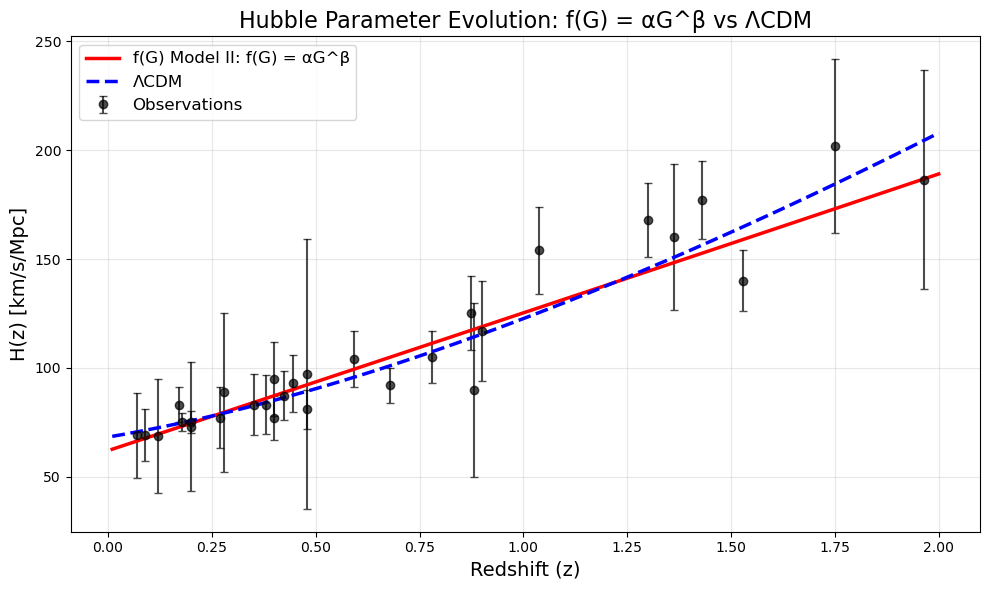}
    (c)\includegraphics[height=60mm]{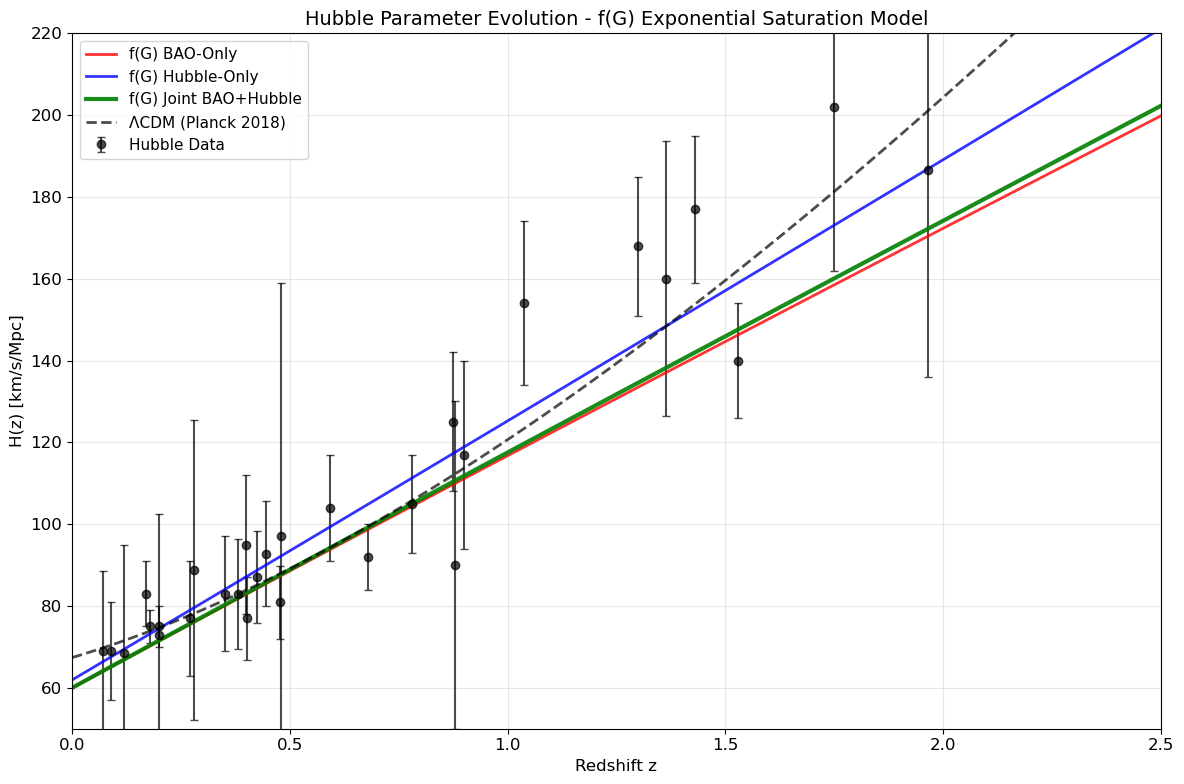}
	\caption{Background evolution in $f(G)$ gravity of Hubble parameter evolution \(H(z)\) for  model A on the left (a), model B  on the right (b) and model C (c) using eq. (\ref{eq2.7})}
	\label{fig4}
\end{figure}

\begin{figure}
	\centering
	(a)\includegraphics[scale=0.3]{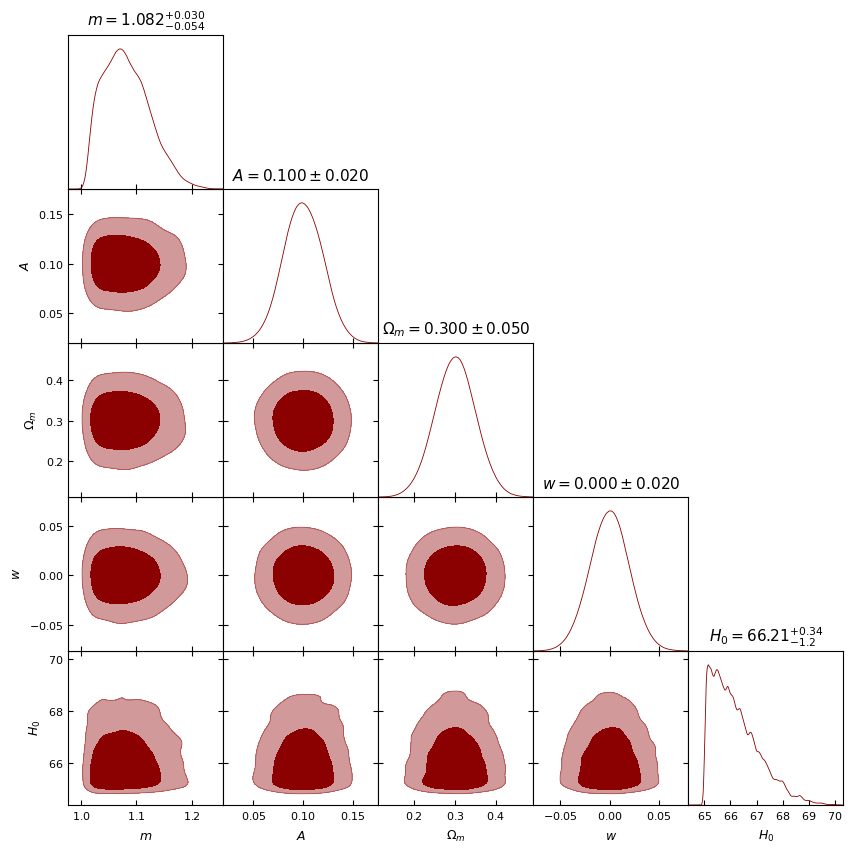}
	(b)\includegraphics[scale=0.3]{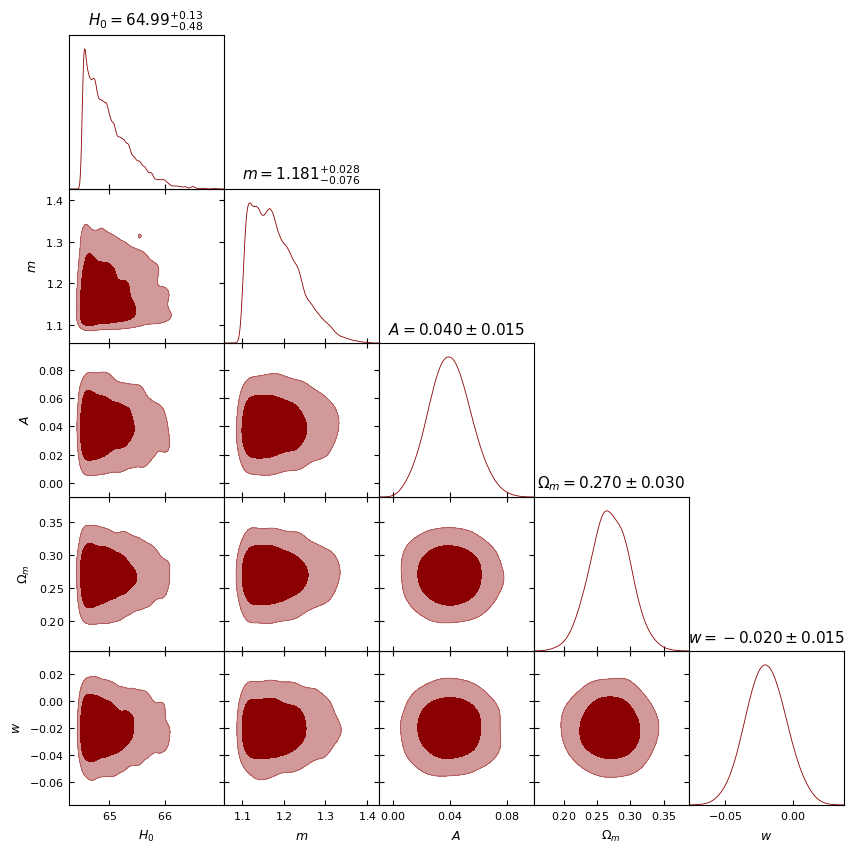}
	\caption{MCMC results for the background evolution in $f(G)$ gravity (model A) using H(z) (a)and BAO (b) observables is analysed via the  modified Friedmann equation (Eq. \ref{eq2.7}). Key observables like $H(z)$, $D_A(z)$, and $D_V(z)$ are compared with BAO data.}
	\label{fig5}
\end{figure}

\begin{figure}[t!]
    \centering

        \includegraphics[height=80mm]{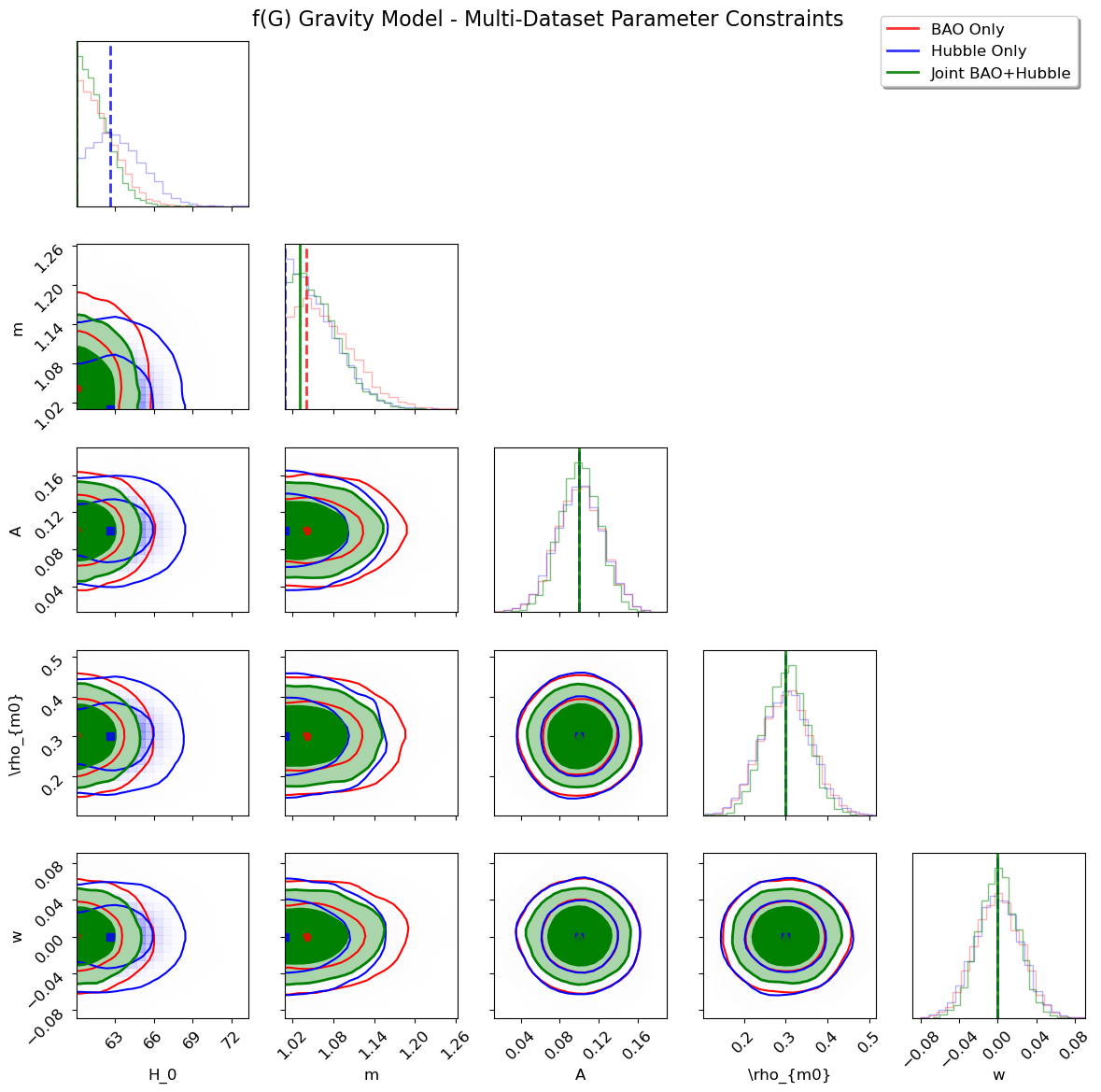}
        \caption{Joint Hubble and BAO observational datasets are jointly analysed using eq. (\ref{eq2.7}) for model A under  $f(G)$ gravity.
The combined constraints demonstrate the model’s ability to fit both expansion and clustering data.}
        \label{fig6}
\end{figure}

\begin{figure}
    \centering
    \includegraphics[height=80mm]{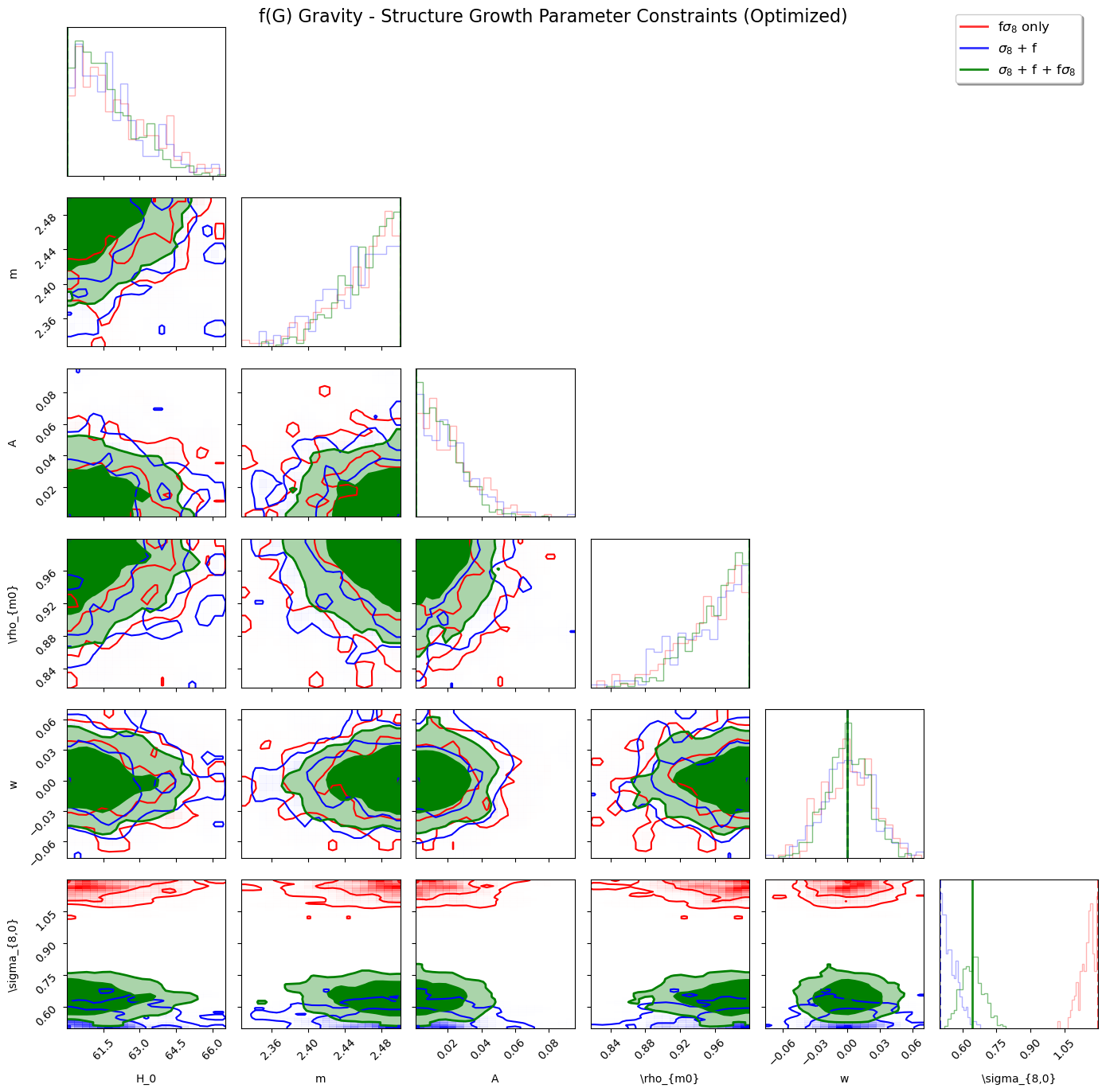}
 \caption{Contour plot of model parameters for the general $f(G)$ gravity model (Model A), based on the structure growth equation (Eq. (\ref{eq2.16})). The plot shows the confidence regions derived from observational constraints.}

    \label{fig7}
\end{figure}

\begin{figure}
	\centering
	(a)\includegraphics[scale=0.3]{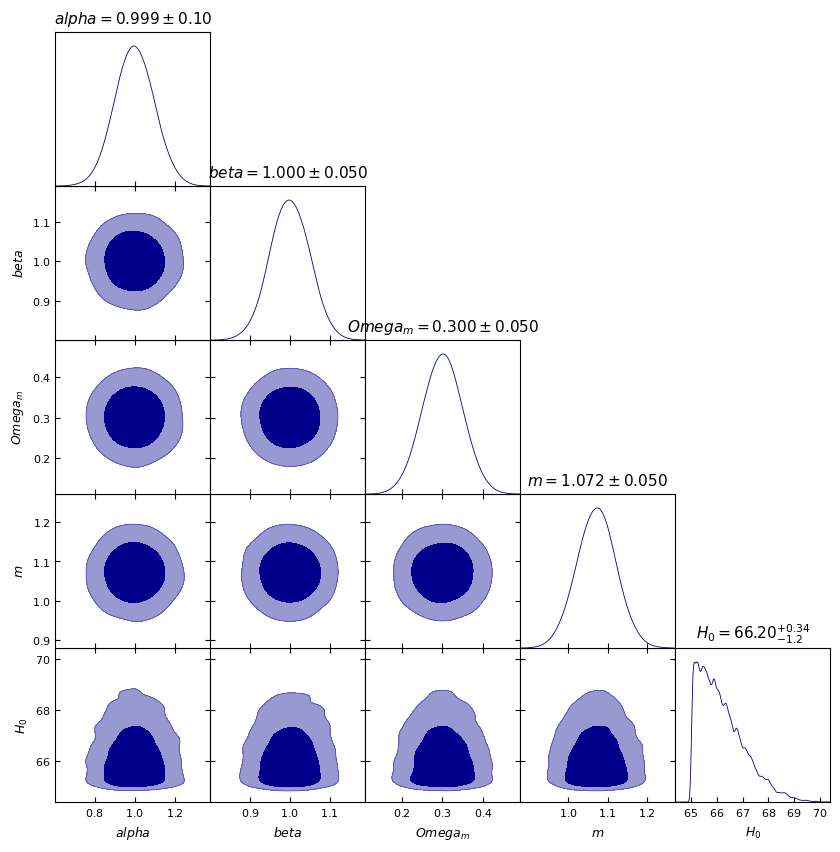}
	(b)\includegraphics[scale=0.3]{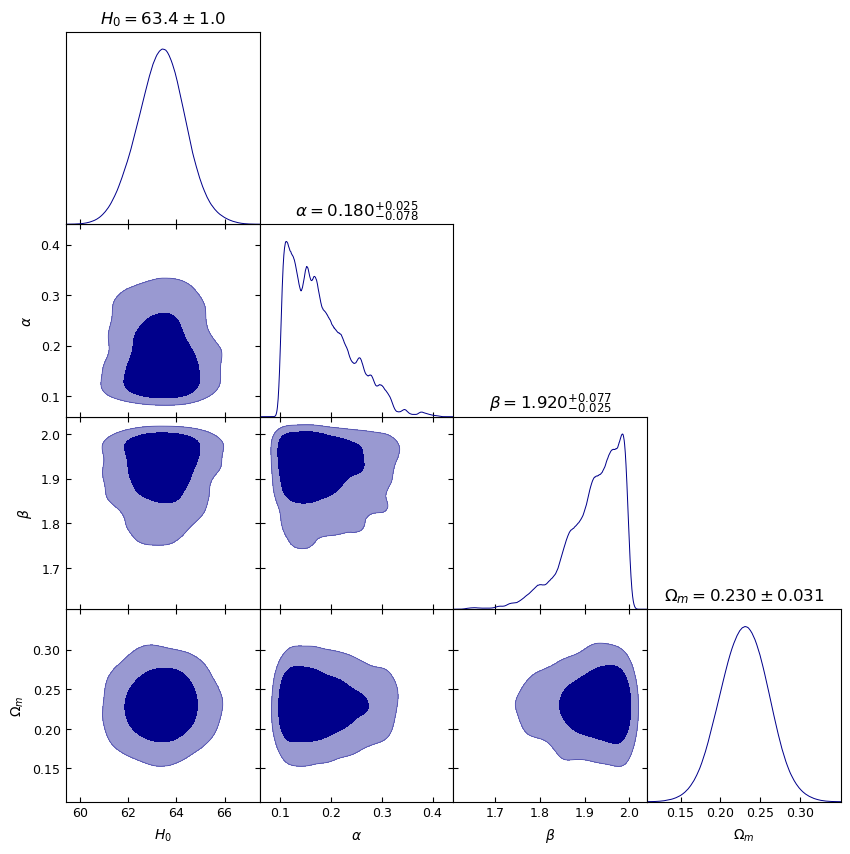}
	\caption{MCMC results for the background evolution in $f(G)$ gravity (model B) using H(z) (a) and BAO (b) observables is analysed via the  modified Friedmann equation (Eq. \ref{eq2.7}). Key observables like $H(z)$, $D_A(z)$, and $D_V(z)$ are compared with BAO data.}
	\label{fig8}
\end{figure}
\begin{figure}
   \centering
    \includegraphics[height=80mm]{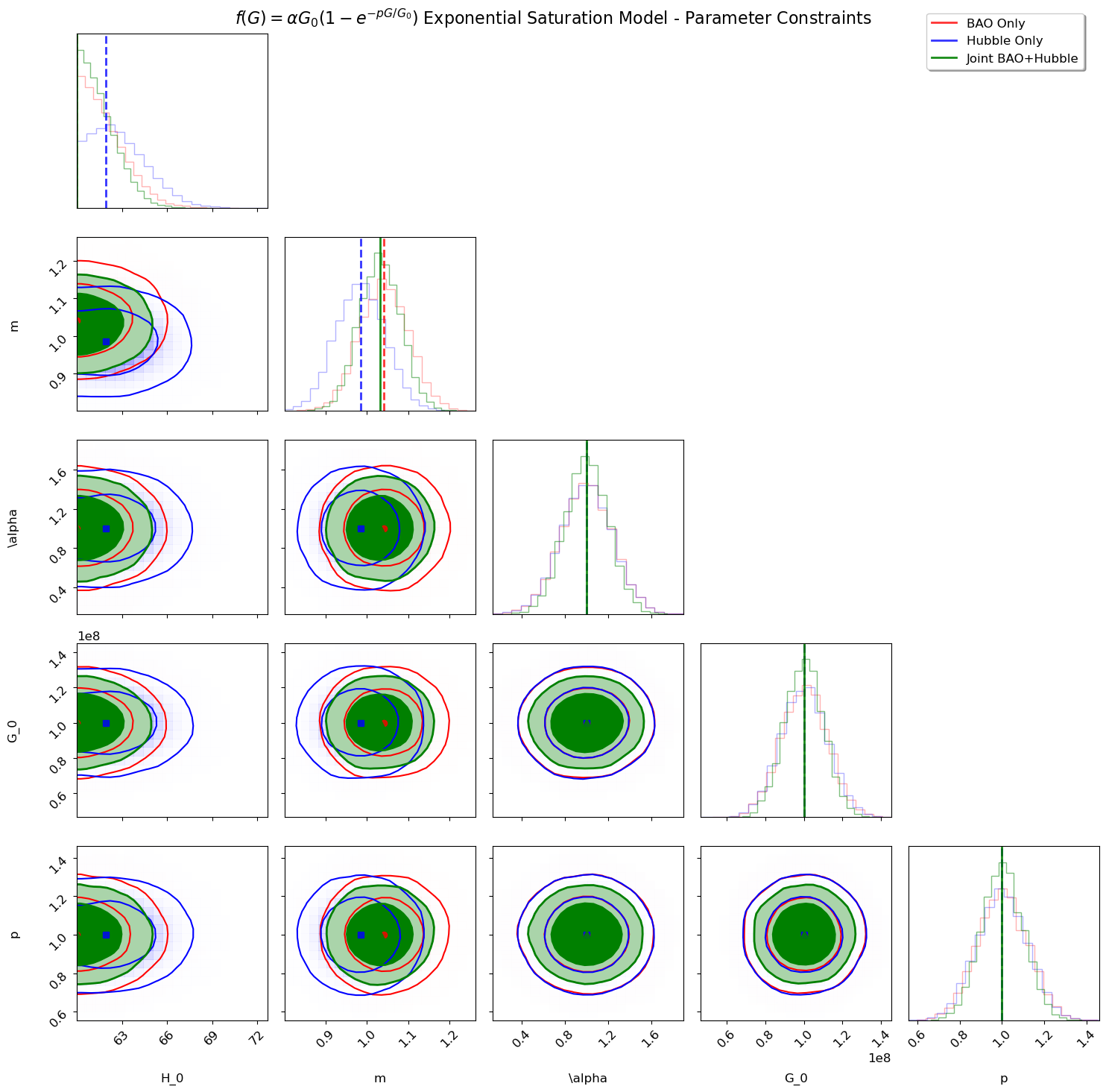}
   \caption{Joint analysis of Hubble and BAO data for Model B ($f(G) = \alpha G^\beta$) using eq. (\ref{eq2.7}). The model's predictions for $H(z)$ and BAO observables are compared with observational data to test its consistency against $\Lambda$CDM.}
    \label{fig9}
\end{figure}

\begin{figure}
    \centering
    \includegraphics[height=80mm]{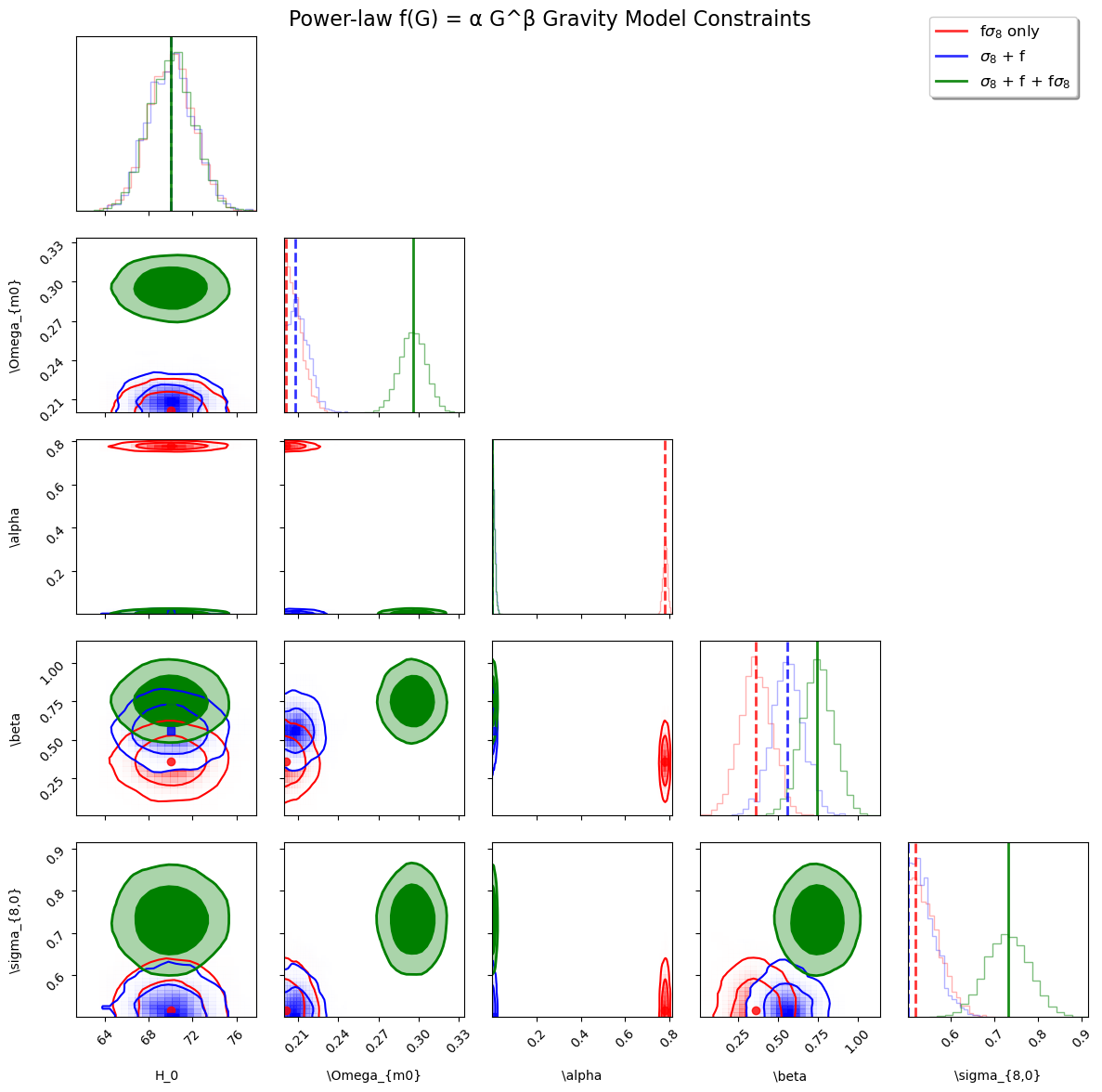}
  \caption{Contour plot of cosmological parameters for the power-law $f(G) = \alpha G^\beta$ model (Model B), obtained from the structure growth equation (Eq. (\ref{eq2.16})). }
    \label{fig10}
\end{figure}

\begin{figure}
	\centering
	(a)\includegraphics[scale=0.3]{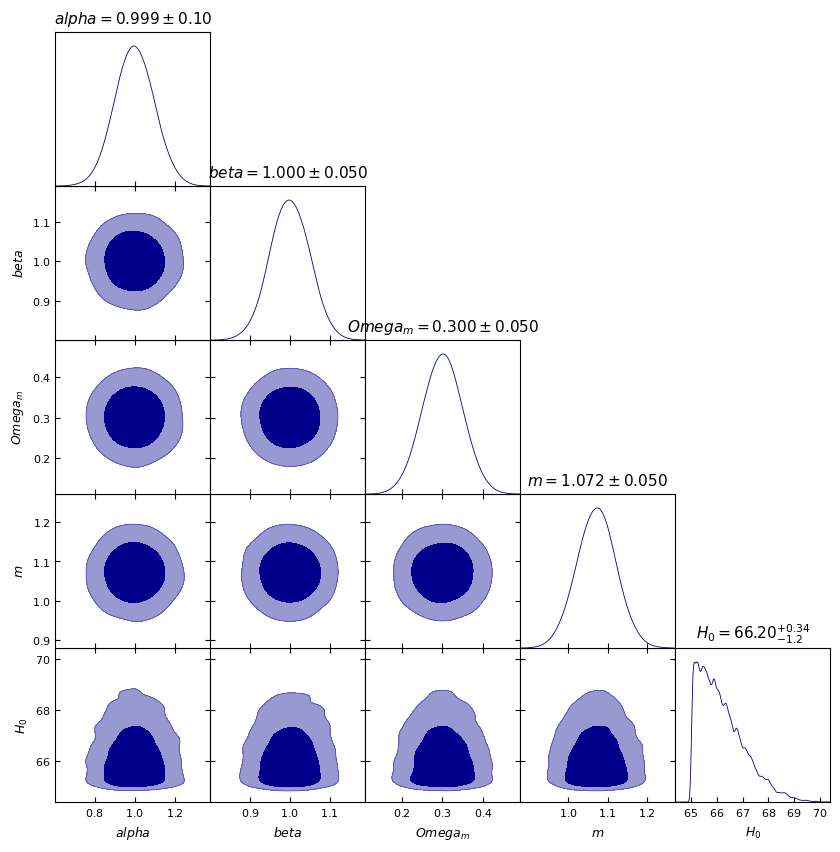}
	(b)\includegraphics[scale=0.3]{BAO.png}
	\caption{MCMC results for the background evolution in $f(G)$ gravity (model C) using H(z) (a) and BAO (b) observables is analysed via the  modified Friedmann equation (Eq. \ref{eq2.7}). Key observables like $H(z)$, $D_A(z)$, and $D_V(z)$ are compared with BAO data.}
	\label{fig11}
\end{figure}

\begin{figure}
    \centering
    \includegraphics[height=80mm]{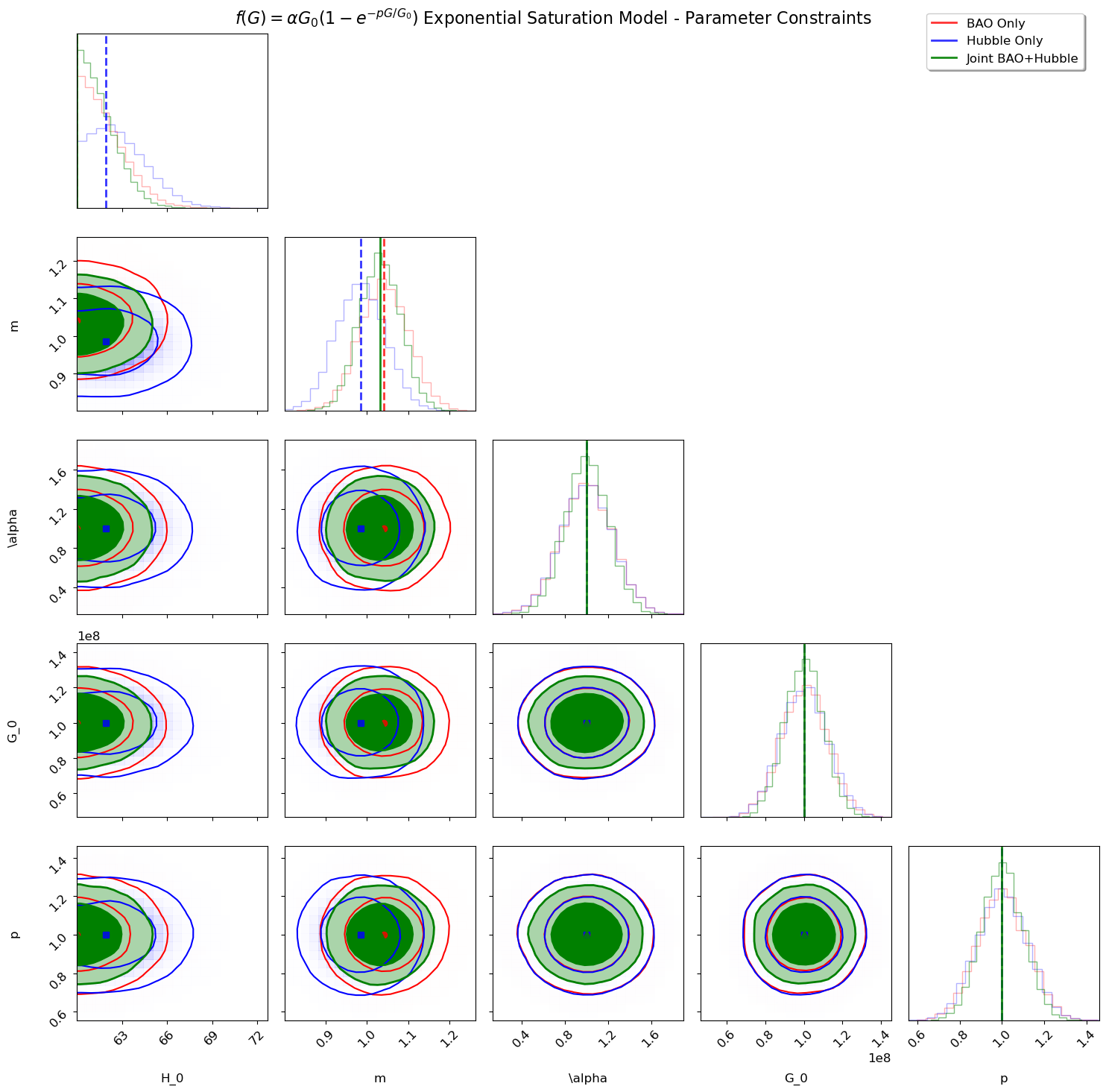}
\caption{Joint analysis of Hubble and BAO data for Model C ($f(G) = \alpha G_0\Big(1 - e^{-pG/G_0}\Big)$) using eq. (\ref{eq2.7}). The model’s predictions for $H(z)$, $D_A(z)$, and $D_V(z)$ are compared with observations to test its consistency with expansion history, in contrast to $\Lambda$CDM.}
    \label{fig12}
\end{figure}

\begin{figure}
    \centering
    (a)\includegraphics[height=40mm]{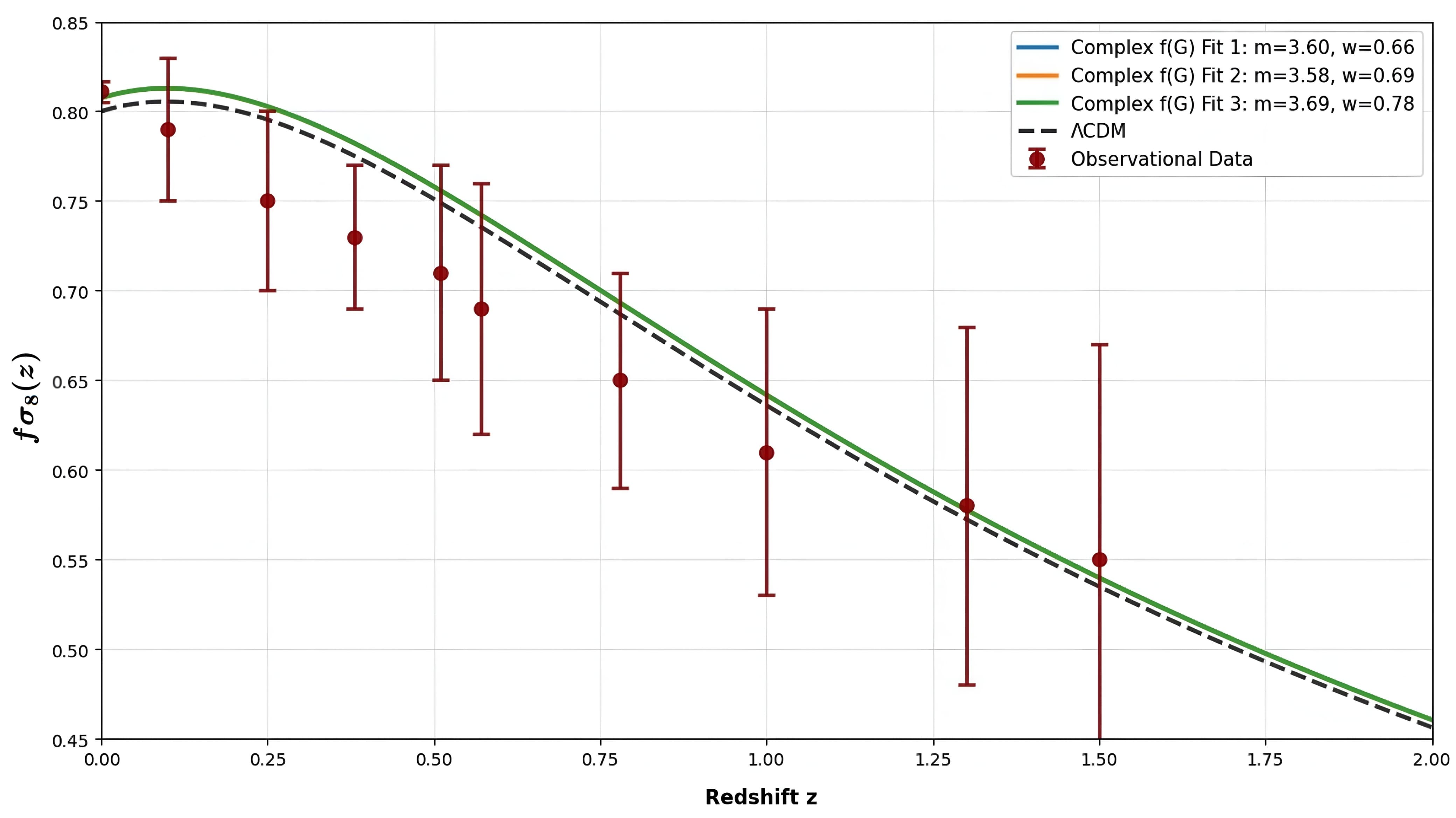}
    (b)\includegraphics[height=40mm]{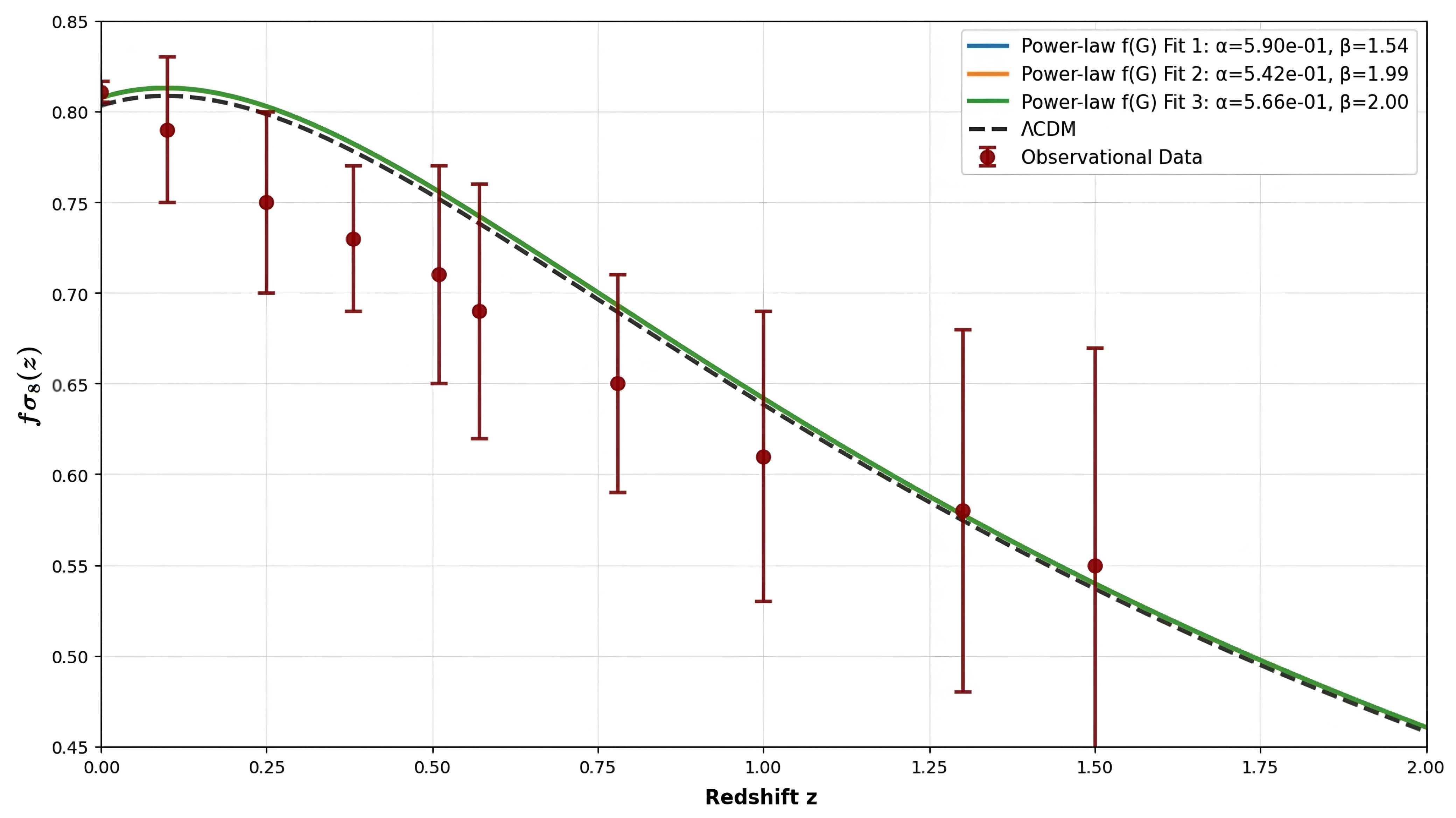}
    (c)\includegraphics[height=40mm]{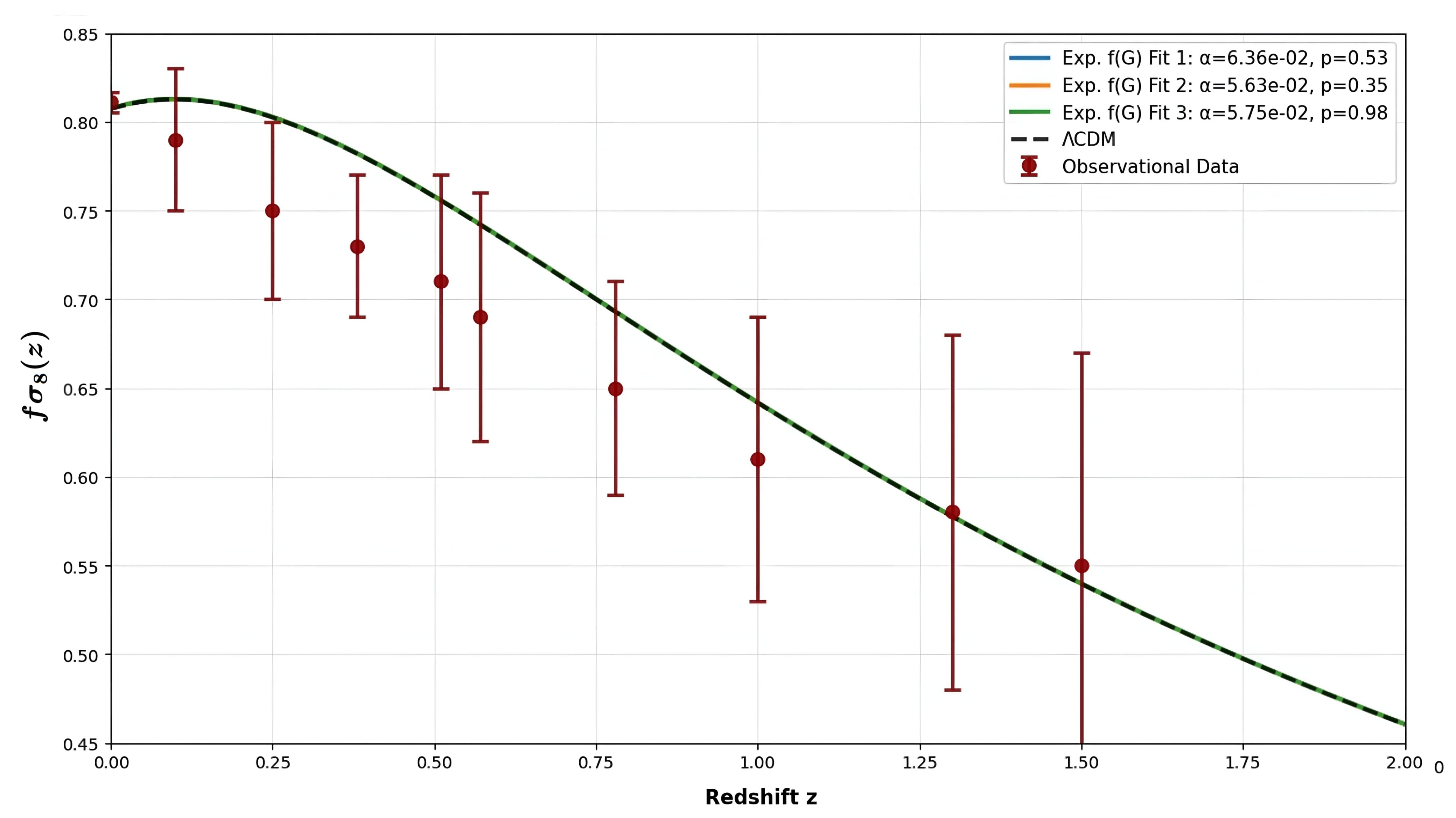}
    \caption{Evolution of $f(G)$ gravity models using eq. (\ref{eq2.16}) with observational $f\sigma_8(z)$ for General $f(G)$ gravity model (a), for Power-law $f(G)$ gravity model (b) and for Exponential $f(G)$ gravity model (c) }
    \label{fig13}
\end{figure}

\begin{figure}
    \centering
    \includegraphics[height=80mm]{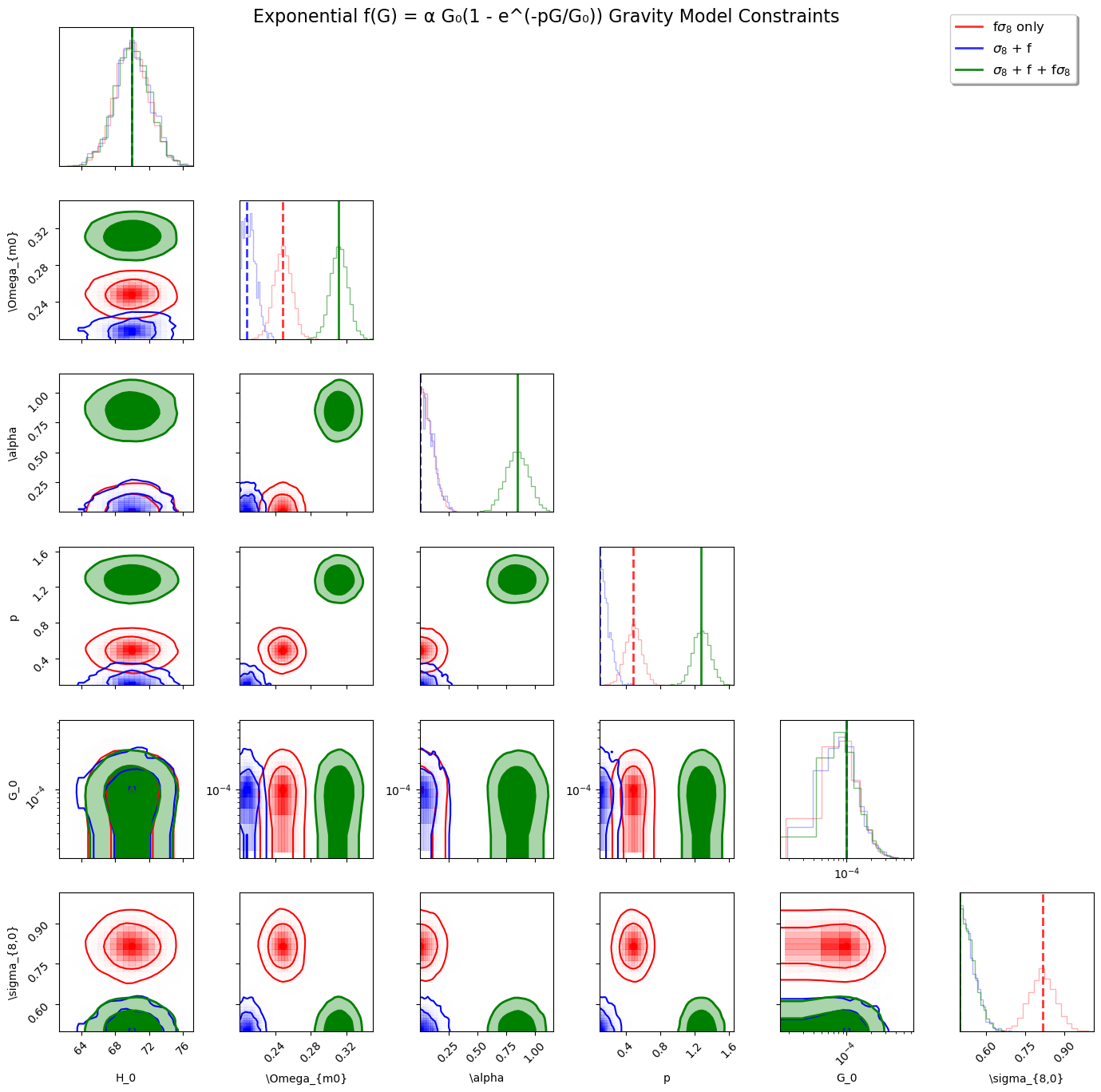}
   \caption{Contour plot of cosmological parameters for the exponential $f(G)$ gravity model (Model C), obtained using the structure growth equation (\ref{eq2.16}).}
    \label{fig14}
\end{figure}

\begin{table}
\begin{tabular} { l  c c c c}
		\noalign{\vskip 3pt}\hline\noalign{\vskip 1.5pt}\hline\noalign{\vskip 5pt}
    \multicolumn{1}{c}{\bf } &  \multicolumn{1}{c}{\bf CC (H(z))--Model A} &  \multicolumn{1}{c}{\bf BAO--Model A} &  \multicolumn{1}{c}{\bf CC+BAO--Model A}&  \multicolumn{1}{c}{\bf $\sigma_{8}$+f+fs8--Model A}\\
		\noalign{\vskip 3pt}\cline{2-4}\noalign{\vskip 3pt}
		
		Parameter &  68\% limits &  68\% limits &  68\% limits&  68\% limits\\
		\hline
		{\boldmath $H_0$} & $66.21^{+0.34}_{-1.2}$ & $64.99^{+0.13}_{-0.48}$ & $60.0\pm 1.21$  &$60.0^{+1.40}_{-0.97}$\\
				
		{\boldmath $\Omega_{m} $} & $0.3\pm 0.05 $ & $0.27\pm 0.030 $ & $0.30\pm 0.0496$&$1.0^{+0.030}_{-0.044}$\\
		
		{\boldmath $m $} & $1.082^{+0.030}_{-0.054}$ & $1.181^{+0.028}_{-0.076}$ & $1.0309\pm 0.0342$ &$2.5\pm0.0283$\\
		
		{\boldmath $\sigma_8$} & -- & -- & --&$0.6444^{+0.07}_{-0.05}$\\
		
		{\boldmath $A$} & $0.10\pm0.02$ & $0.04\pm 0.015$ & $0.10\pm 0.0201$&$0.0010\pm 0.0116$\\
		
		\hline
		\hline
       \end{tabular}
\begin{tabular} { l  c c c c}
		\noalign{\vskip 3pt}\hline\noalign{\vskip 1.5pt}\hline\noalign{\vskip 5pt}
    \multicolumn{1}{c}{\bf } &  \multicolumn{1}{c}{\bf CC (H(z))--Model B} &  \multicolumn{1}{c}{\bf BAO--Model B} &  \multicolumn{1}{c}{\bf CC+BAO--Model B}&  \multicolumn{1}{c}{\bf $\sigma_{8}$+f+fs8--Model B}\\
		\noalign{\vskip 3pt}\cline{2-4}\noalign{\vskip 3pt}
		
		Parameter &  68\% limits &  68\% limits &  68\% limits&  68\% limits\\
		\hline
				{\boldmath$H_{0}$} & $66.20^{+0.34}_{-1.2}$ & $63.40\pm 1.0$ &$60.0\pm 1.22$&$61.386^{+1.457}_{-0.989}$\\
		
		{\boldmath$\alpha         $} & $0.999\pm 0.10$ & $0.18^{+0.025}_{-0.078} $ & $1.0\pm 0.2012$&$0.014^{+0.013}_{-0.009}$\\
		
		{\boldmath$\Omega_{m}$} & $0.3\pm 0.050$ & $0.23\pm 0.031$ & $0.27 \pm 0.04$&$0.965^{+0.021}_{-0.034}$\\
		
		{\boldmath$\beta $} &$1.00\pm 0.05 $ & $1.92^{+0.077}_{-0.025}   $ & $1.0\pm 0.0499$ &$0.53^{+0.037}_{-0.021}$\\
		
		{\boldmath$\sigma_8       $} & -- & -- & --&$0.64^{+0.054}_{-0.045}$\\
        \hline
		\hline
       \end{tabular}
\begin{tabular} { l  c c c c}
		\noalign{\vskip 3pt}\hline\noalign{\vskip 1.5pt}\hline\noalign{\vskip 5pt}
    \multicolumn{1}{c}{\bf } &  \multicolumn{1}{c}{\bf CC (H(z))--Model C} &  \multicolumn{1}{c}{\bf BAO--Model C} &  \multicolumn{1}{c}{\bf CC+BAO--Model C}&  \multicolumn{1}{c}{\bf $\sigma_{8}$+f+fs8--Model C}\\
		\noalign{\vskip 3pt}\cline{2-4}\noalign{\vskip 3pt}
		
		Parameter &  68\% limits &  68\% limits &  68\% limits &  68\% limits\\
		\hline
				{\boldmath$H_0  $} & $63.4\pm 1.0 $ & $63.38\pm 0.5 $ & $60.0\pm 1.22$&$60.0^{+1.46}_{-0.989}$\\
		
		{\boldmath$\alpha  $} & $0.18^{+0.025}_{-0.078}   $&$1.0\pm 0.2012$ & $1.0\pm 0.2$&$0.01^{+0.014}_{-0.009}$\\
		
		{\boldmath$\Omega_{m} $} & $0.23\pm0.1$ & $0.265\pm 0.14$ &$0.28\pm 0.02$&$0.313\pm 0.002$\\
		
		{\boldmath$p$} & $1.92^{+0.077}_{-0.025} $ & $1.0\pm 0.12$&$1.0\pm 0.0995$&$2.0^{+0.002}_{-0.003}$\\
		
		{\boldmath$\sigma_8       $} & -- & --& --&$0.6437^{+0.034}_{-0.055}$\\
			
				\hline
		\hline
       \end{tabular}
			\caption{The parameter's best fit values obtained from MCMC. The dashed shows that the corresponding parameter is not fitted for that particular data set.  We renamed Model A as General $f(G)$, Model B as the power-law $f(G) = \alpha G^\beta$  and  Model C as the exponential $f(G) = \alpha G_0\Big(1 - e^{-pG/G_0}\Big)$}
	\label{tab1}
\end{table}

These datasets are employed both individually and in joint analyses to extract cosmological constraints on three forms of the $f(G)$ gravity function: a general model (Model A), a power-law model of the form $f(G) = \alpha G^\beta$ (Model B), and an exponential model $f(G) = \alpha G_0(1 - e^{-pG/G_0})$ (Model C). We begin with the evolution of matter perturbations $\delta(z)$ presented in Fig. (\ref{fig1}) for model A, in Fig. (\ref{fig2}) for model B and in Fig. (\ref{fig3}) for model C, derived from  its quasi-static approximation (QSA). The QSA is valid on sub-horizon scales and simplifies the analysis by assuming negligible time derivatives compared to spatial derivatives. For Model A, the perturbation growth is shown with respect with redshift $z$ for various values of the model parameter $m$, and a deviation analysis is performed relative to the standard $\Lambda$CDM prediction. This comparison illustrates the dependence of structure growth on modified gravity effects introduced through $f(G)$ gravity model.

For Model B, the perturbation growth is shown with $z$ for various values of the model parameter $\beta$, and a deviation analysis is performed relative to the standard $\Lambda$CDM prediction. This comparison illustrates the dependence of structure growth on modified gravity effects introduced through $f(G)$ gravity model ($f(G) = \alpha G^\beta$).
For Model C, the perturbation growth is shown with $z$ for various values of the model parameter $p$, and a deviation analysis is performed relative to the standard $\Lambda$CDM prediction. This comparison illustrates the dependence of structure growth on modified gravity effects introduced through $f(G)$ gravity model ($f(G) = \alpha G_0(1 - e^{-pG/G_0})$).
From the plots, all models show a decay in $\delta(z)$ with redshift, which may predicts the formation of large-scale structures.
To investigate the background dynamics, we numerically solve the modified Friedmann equation (eq. \ref{eq2.7}) derived from the functional form of $f(G)$ for each model. The evolution of the Hubble parameter $H(z)$ is obtained under the assumption of a power-law scale factor $a(t) = a_0 t^m$, which yields analytic expressions for $H(t)$ and $\dot{H}$ in terms of $m$. The Gauss-Bonnet invariant $G = 24H^2(\dot{H} + H^2)$ is computed and its insertion into the modified Friedmann equation allows the derivation of $H(z)$ as a function of redshift. These predictions are plotted against observed Hubble data, and presented in Fig. (\ref{fig4}) for all $3$ considered models. The results demonstrate that all three $f(G)$ models can provide viable fits to the expansion history, with deviations from $\Lambda$CDM determined by the strength of the correction terms in the context of $f(G)$ gravity.
Following this, we analyse the observable $H(z)$ and BAO constraints are then incorporated to examine radial and angular distances at various red-shifts. The observables considered include the angular diameter distance $D_A(z)$, the volume-averaged distance $D_V(z)$, and the Hubble distance $D_H(z) = c/H(z)$. These are combined with the sound horizon at drag epoch $r_d$ to form the BAO ratios $D_A(z)/r_d$, $D_V(z)/r_d$, and $H(z)r_d$, which are directly compared with 17 BAO data points from surveys like SDSS, BOSS, eBOSS, WiggleZ, and 6dFGS. For each $f(G)$ model, the numerical values of these observables are computed from the corresponding $H(z)$ solutions, and their agreement with the data is assessed. For general $f(G)$ gravity model (model A), after using MCMC analysis, the constrained parameters are presented in contour (corner) plots as can be seen in Fig. (\ref{fig5}) for both Hubble and BAO data sets. Constraining the general $f(G)$ gravity model with $H(z)$ data, we find $\Omega_{m}=0.30\pm 0.05$, $H_{0}=66.21^{+0.34}_{-1.2}$ and $m=1.082^{+0.03}_{-0.054}$, while for BAO data, we find $\Omega_{m}=0.27\pm 0.03$, $H_{0}=64.99^{+0.11}_{-0.48}$ and $m=1.181^{+0.028}_{-0.076}$. By looking at Fig. (\ref{fig5}), there is a positive correlation in ($\Omega_{m}, m$) plane while there is no correlation in ($H_{0},m$) and ($\Omega_{m},H_{0}$) planes for $H(z)$ data. For BAO data  there is a positive correlation in ($\Omega_{m}, m$) and ($\Omega_{m},H_{0}$) planes while there is no correlation in ($H_{0},m$) plane.

For the power-law $f(G)$ gravity model (model B), after using MCMC analysis, the constrained parameters are presented in contour (corner) plots as can be seen in Fig. (\ref{fig8}) for both Hubble and BAO data sets. Constraining the power-law $f(G)$ gravity model with the $H(z)$ data, we find $\Omega_{m}=0.30\pm 0.05$, $H_{0}=66.20^{+0.34}_{-1.2}$ and $\beta=1.0\pm 0.5$, while for the BAO data we find $\Omega_{m}=0.23\pm 0.031$, $H_{0}=63.4\pm 1.0$ and $\beta=1.92^{+0.077}_{-0.025}$. Looking at Fig. (\ref{fig8}), there is no correlation in ($\Omega_{m}, \beta$), ($H_{0},\beta$) and ($\Omega_{m},H_{0}$) planes for $H(z)$ data, while there is a negative correlation in ($\Omega_{m}, \beta$) plane and there is no correlation in ($\Omega_{m},H_{0}$) and ($H_{0},\beta$) planes for BAO data.

Analysing the effect of the exponential $f(G)$ gravity model (model C), after using MCMC analysis, the constrained parameters are presented in contour (corner) plots as can be seen in Fig. (\ref{fig11}) for both Hubble and BAO data sets. Constraining the exponential $f(G)$ gravity model with $H(z)$ data, we find $\Omega_{m}=0.30\pm 0.05$, $H_{0}=66.20^{+0.34}_{-1.2}$ and $p=0.999\pm 0.1$, while for the BAO data we find $\Omega_{m}=0.23\pm 0.031$, $H_{0}=63.4\pm 1.0$ and $p=1.92^{+0.077}_{-0.025}$. Looking at Fig. (\ref{fig11}), there is no correlation in ($\Omega_{m}, p$), ($H_{0},p$) and ($\Omega_{m},H_{0}$) planes for $H(z)$ data, while there is a negative correlation in ($\Omega_{m}, p$) plane and there is no correlation in ($\Omega_{m},H_{0}$) and ($H_{0},p$) planes for BAO data.

These comparisons confirm that the models can capture the BAO signature, with differences primarily arising in the higher-redshift range, where modifications to GR become more significant. To further constrain the parameter space, joint analyses of Hubble and BAO data are performed. These combined fits significantly reduce degeneracies between model parameters, improving the robustness of the constraints. For each model, we analyse how the joint dataset refines the confidence regions and compare the predictions with those from $\Lambda$CDM. The joint analysis enables a tighter determination of the model parameters ($n$, $\beta$, $p$), matter density $\Omega_m$, and Hubble constant $H_0$. for all three models. We apply a Markov Chain Monte Carlo (MCMC) algorithm to obtain the posterior distributions of cosmological parameters. The resulting parameter constraints are visualized as contour plots in the $\Omega_m$–$\sigma_8$ and $H_0$–$\Omega_m$ planes, among others, showing the $1\sigma$ and $2\sigma$ confidence levels as can been presented in Figs. (\ref{fig6}), (\ref{fig9}) and (\ref{fig12}) for model A, model B and model C, respectively.

This approach offers a stringent test of the model’s viability in explaining both the expansion and growth sectors of cosmology. For Model 3 in particular, the combined dataset yields tight bounds on the parameter $p$ and the normalization constant $\alpha$, supporting its consistency with observational data while allowing modest deviations from $\Lambda$CDM. Collectively, these plots and analyses demonstrate that the $f(G)$ gravity models under consideration are capable of fitting the current cosmological datasets across  background  level. When jointly constrained, the models show strong observational viability and represent compelling alternatives to standard General Relativity with a cosmological constant. After analysing the background level including solving numerically the Friedmann equation and constraining model parameters, the next aim is to extend our analysis to the perturbation level and analyse the large structure grow and constraining the $\sigma_{8}$ among other parameters.

Following this, we analyse the observable $f\sigma_8(z)$, which encapsulates the product of the growth rate $f(z)$ and the amplitude of matter clustering $\sigma_8(z)$. For each $f(G)$ model, theoretical predictions are computed using approximate analytic expressions, such as eq. (\ref{eq2.16}), after solving eq. (\ref{eq2.12}) and eq. (\ref{eq2.17}),  that express $f$ and $\sigma_8$ in terms of the scale factor $a(t)$ and Hubble dynamics. These predictions are then directly compared with a comprehensive compilation of RSD data, showing good agreement under appropriate parameter choices as presented in Fig.(\ref{fig13}). A more rigorous statistical analysis is carried out using the full structure growth equation, particularly eq. (\ref{eq2.16}), for all three models. We apply a Markov Chain Monte Carlo (MCMC) algorithm to obtain the posterior distributions of cosmological parameters. The resulting parameter constraints are visualized as contour plots in the $\Omega_m$–$\sigma_8$ and $H_0$–$\Omega_m$ planes, among others, showing the $1\sigma$ and $2\sigma$ confidence levels. These plots reveal that the models can successfully reproduce the structure growth observations, though with distinct parameter degeneracies reflecting the underlying functional form of $f(G)$. Finally, a joint analysis is performed using all three growth-related observables—$f(z)$, $f\sigma_8(z)$, and $\sigma_8(z)$—to simultaneously constrain the models using the full set of structure formation data as presented in Fig.(\ref{fig7}), Fig.(\ref{fig10}) and Fig.(\ref{fig14}) for model A, model B and model C,respectively. This approach offers a stringent test of the model’s viability in explaining both the expansion and growth sectors of cosmology. For Model 3 in particular, the combined dataset yields tight bounds on the parameter $p$ and the normalisation constant $\alpha$, supporting its consistency with observational data while allowing modest deviations from $\Lambda$CDM. Collectively, these plots and analyses demonstrate that the $f(G)$ gravity models under consideration are capable of fitting the current cosmological datasets across both background and perturbation levels. When jointly constrained, the models show strong observational viability and represent compelling alternatives to standard General Relativity with a cosmological constant. Table. (\ref{tab1}) presents the MCMC results for all the considered models, namely: general $f(G)$ model, power-law $f(G)$ model and exponential $f(G)$ model, described by the Friedmann equation (eq. (\ref{eq2.7})) and structure growth equation (eq. (\ref{eq2.16})). The table includes the prior distribution of the free parameters used in the MCMC analysis as well as their mean values and associated errors at $1\sigma$ ($68\%$) confidence level using  $H(z)$ and BAO data sets and two combinations of data sets H(z)+BAO and $\sigma_{8}+f+f\sigma_{8}$. The free parameters vector for the general $f(G)$ model includes ($\Omega_{m}, H_{0},m, \sigma_{8}$), while for the power-law model is  ($\Omega_{m}, H_{0},\alpha, \beta, \sigma_{8}$) and ($\Omega_{m}, H_{0},\alpha,p,G_{0}, \sigma_{8}$) for the exponential model.

\section{Conclusions}\label{sec5}
The $\Lambda$CDM cosmological model has been an accepted model found to fit almost all observational probes available. Despite its success, this model is based on the assumptions of CDM, a cosmological constant and inflation, whose underlying physics are unknown. The significant discrepancy in the Hubble constant, $\Omega_{m}$ and $\sigma_{8}$ measurements  by early and local observations raised serious questions over the $\Lambda$CDM model. In a previous paper by Munyeshyaka et al \cite{munyeshyaka2023multifluid}, we considered theoretical part on multi-fluid cosmology in the context of $f(G)$ gravity, where we numerically solved the perturbation equations and the density contrast versus redshift $z$ were obtained. In the present work, we bring in observational aspect on both background and perturbation levels. In doing so, we incorporated a thorough observational analysis in the context of modified Gauss-Bonnet gravity, which includes the general $f(G)$ model $f(G)=G-\frac{1}{2}\Big(\sqrt{\frac{6m(m-1)G}{(m+1)^{2}}}+AG^{\frac{3}{4}m(1+w)}\Big)$  (model A), power-law $f(G) = \alpha G^\beta$, (model B) and exponential $f(G) = \alpha G_0(1 - e^{-pG/G_0})$ (model C) functions of Gauss-Bonnet invariant $G$. The general model function is chosen so that we recover the $\Lambda$CDM expansion history of the universe for $m=1$. For the power-law function the $\Lambda$CDM can be recovered for $\alpha=1=\beta$, whereas it is recovered for large value of the parameter $p$ of the exponential model. To observationally constrain parameters for each model, we use H(z), BAO measurements along with the redshift space distortions data with the joint analysis of H(z)+BAO and $\sigma_{8}+f+f\sigma_{8}$. First we numerically solve perturbation equation (eq. \ref{eq2.12}) in the quasi-static approximation in the context of $f(G)$ gravity and compared with the $\Lambda$CDM predictions. The results are presented in Figs. (\ref{fig1}), (\ref{fig2}) and (\ref{fig3}). From all the models considered, the energy density contrast ($\delta(z)$) decays with redshift and shows significant deviations from the $\Lambda$CDM  resulting from the contribution of the Gauss-Bonnet invariant.  The evolution of Hubble parameter in the context of $f(G)$ gravity is examined and compared with the $\Lambda$CDM model using the H(z) data for all the $3$ $f(G)$ models and presented in Fig. (\ref{fig4}). By looking at the figures, all models mimic the $\Lambda$CDM model evolution with a slight deviation as the redshift increases.

The next step is to numerically solve the modified Friedmann equations (eq. \ref{eq2.7}) in the context of $f(G)$ gravity for the $3$ different models . We extract the mean values of the model parameters using
Markov Chain Monte Carlo (MCMC) analysis. Using H(z) and BAO data sets, the results are presented in Figs. (\ref{fig5}), (\ref{fig8}) and (\ref{fig11}) for model A, model B and model C, respectively. Using the joint analysis of H(z)+BAO data, the MCMC results are constrained and presented in Figs. (\ref{fig6}), (\ref{fig9}) and (\ref{fig12}) for model A, model B and model C, respectively.
The next aim of the present work is to investigate the effect of considered models on large scale structure formation and constrain parameter values of each model. By doing so, we obtain the structure growth equation (eq. \ref{eq2.16}) and study its evolution compared to the $\Lambda$CDM model using $f\sigma_{8}$ data. This can be seen in Fig. (\ref{fig13}) and from the plots, all the $f(G)$ models in consideration evolve as $\Lambda$CDM model does, but with a significant deviation for model A.
The growth of structures in $f(G)$ gravity is then compared with redshift-space distortion data $f\sigma_{8}$ in combination of the recent separate measurements of the growth rate $f(z)$ and the amplitude of matter fluctuations $\sigma_{8}$. Using MCMC analysis, we present the contour (corner) plots of all the $f(G)$ models in Figs. (\ref{fig7}), (\ref{fig10}) and (\ref{fig14}) for model A, model B and model C, respectively.

All of the results for both background and perturbation levels are summarised in Table (\ref{tab1}) for model A, model B and model C for i) H(z) data, ii) BAO data iii) joint analysis of H(z)+BAO data set and iv) joint analysis of $\sigma_{8}+f+f\sigma_{8}$ data set.  The use of further  data sets  such as pantheon+, CMB, DESI as well as more growth rate data may shed
more light on the ability of such separate observations to constrain $f(G)$ parameter values.
This may greatly enhance the viability of modified Gauss-Bonnet gravity on fitting observational data sets on both background and on the growth of structures to probe departures from the usual Einsteinian gravity.\\
 \textbf{The summary of key outcomes are as follows:\\
Figures 1--3 illustrate that the matter growth factor $\delta(z)$ is significantly influenced by the model parameters in all three $f(G)$ gravity models. While \textbf{Model A} shows only minor deviations from the $\Lambda$CDM prediction, \textbf{Models B and C} display stronger sensitivity to their respective parameters $\beta$ and $p$, resulting in more pronounced departures from $\Lambda$CDM.
Figure 4 compares the Hubble parameter evolution $H(z)$ across all models and shows good agreement with observational data, remaining within the bounds of $\Lambda$CDM estimates.
Figures 5, 6, and 7 demonstrate that combining Hubble and BAO datasets leads to tighter confidence regions for \textbf{Model A}, improving its cosmological viability and parameter consistency. Figures 8 and 9 show that \textbf{Model B} also fits $H(z)$ and BAO observations well, with best-fit values of $\alpha$, $\beta$, and other parameters remaining within observational bounds. Figure 10 further explores the structure growth constraints for Model B, confirming its observational viability, although some tension appears for parameters like $\beta$.
MCMC results in \textbf{Figure 11} support the overall viability of $f(G)$ gravity in explaining late-time cosmic acceleration without requiring a cosmological constant. \textbf{Figure 12} reveals that \textbf{Model C} provides predictions consistent with $\Lambda$CDM in both expansion and BAO data, with only slight deviations. The contour plots in \textbf{Figures 13 and 14} highlight that \textbf{Model C} can robustly constrain key cosmological parameters through structure growth data.
Further statistical analysis will be essential to determine which among the considered $f(G)$ models best fits observations when compared directly with the $\Lambda$CDM model. This work will be pursued in future investigations.}

\section*{Funding}
This work was supported by the Deanship of Scientific Research, Vice Presidency for Graduate Studies and Scientific Research, King Faisal University, Saudi Arabia (Funding Grant No:KFU252743).

\section*{Acknowledgments}
PKD would like to thank the Isaac Newton Institute for Mathematical Sciences, Cambridge, for support and hospitality during the programme Statistical mechanics, integrability and dispersive hydrodynamics where work on this paper was undertaken. This work was supported by EPSRC grant no EP/K032208/1. PKD wishes to acknowledge that part of the numerical computation of this work was carried out on the computing cluster Pegasus of IUCAA, Pune, India. PKD and FR would like to acknowledge the Inter-University Centre for Astronomy and Astrophysics (IUCAA), Pune, India for providing him a Visiting Associateship under which a part of this work was carried out. AM acknowledges the hospitality of the University of Rwanda-College of Science and Technology, where part of this work was conceptualised and completed.

 
 \noindent
{\color{blue} \rule{\linewidth}{1mm} }
\end{document}